%% file: main.tex
\documentclass[acmtog, authorversion]{acmart}
\usepackage{booktabs} % For formal tables
\usepackage{times}
\usepackage{epsfig}
\usepackage{graphicx}
\usepackage{amsmath}
\usepackage{amssymb}
\usepackage{multirow}
\usepackage{comment}
\usepackage{enumitem}
\usepackage{subfigure}
\usepackage{wrapfig}

\setcitestyle{square} % nosort to allow for manual chronological ordering

\usepackage[ruled]{algorithm2e} % For algorithms

\SetAlFnt{\small}
\SetAlCapFnt{\small}
\SetAlCapNameFnt{\small}
\SetAlCapHSkip{0pt}

\input{symbols_format.tex}

\AtBeginDocument{%
  \providecommand\BibTeX{{%
    \normalfont B\kern-0.5em{\scshape i\kern-0.25em b}\kern-0.8em\TeX}}}
    
\acmJournal{TOG}
\acmYear{2020}
\acmVolume{39}
\acmNumber{4}
\acmMonth{7}

\setcopyright{acmcopyright}

\acmDOI{10.1145/3386569.3392379}
\makeatletter
\def\@copyrightspace{\relax}
\let\@authorsaddresses\@empty
\makeatother

\begin{document}

% Title portion
\title{RigNet: Neural Rigging for Articulated Characters}

% DO NOT ENTER AUTHOR INFORMATION FOR ANONYMOUS TECHNICAL PAPER SUBMISSIONS TO SIGGRAPH 2019!
\author{Zhan Xu}
\author{Yang Zhou}
\author{Evangelos Kalogerakis}
\affiliation{%
\institution{University of Massachusetts Amherst}
}
\email{ {zhanxu,yangzhou,kalo}@cs.umass.edu}

\author{Chris Landreth}
\author{Karan Singh}
\affiliation{%
\institution{University of Toronto}
}
\email{{chrisl,karan}@dgp.toronto.edu}

%\renewcommand\shortauthors{Xu, Z. et al}

\begin{abstract}
We present {\it RigNet}, an end-to-end automated method for producing animation rigs from input character models.
Given an input 3D model representing an articulated character, {\it RigNet} predicts a skeleton that matches the animator expectations in joint  placement and topology. It also estimates surface skin weights based on the predicted skeleton. Our method is based on a deep architecture that directly operates on the mesh representation without making assumptions on shape class and structure. The architecture is trained on a large and diverse collection of rigged models, including their mesh, skeletons and corresponding skin weights.
Our evaluation is three-fold: we show  better results than prior art when quantitatively compared to animator rigs; qualitatively we show that our rigs  can be expressively posed and animated at multiple levels of detail; and finally, we evaluate the impact of various algorithm choices  on our output rigs. 
\footnote{
Our project page with source code, datasets, and supplementary video is available at
\textcolor{blue}{https://zhan-xu.github.io/rig-net  }
}
\end{abstract}



\begin{teaserfigure}
\centering
\includegraphics[width=1.0\textwidth]{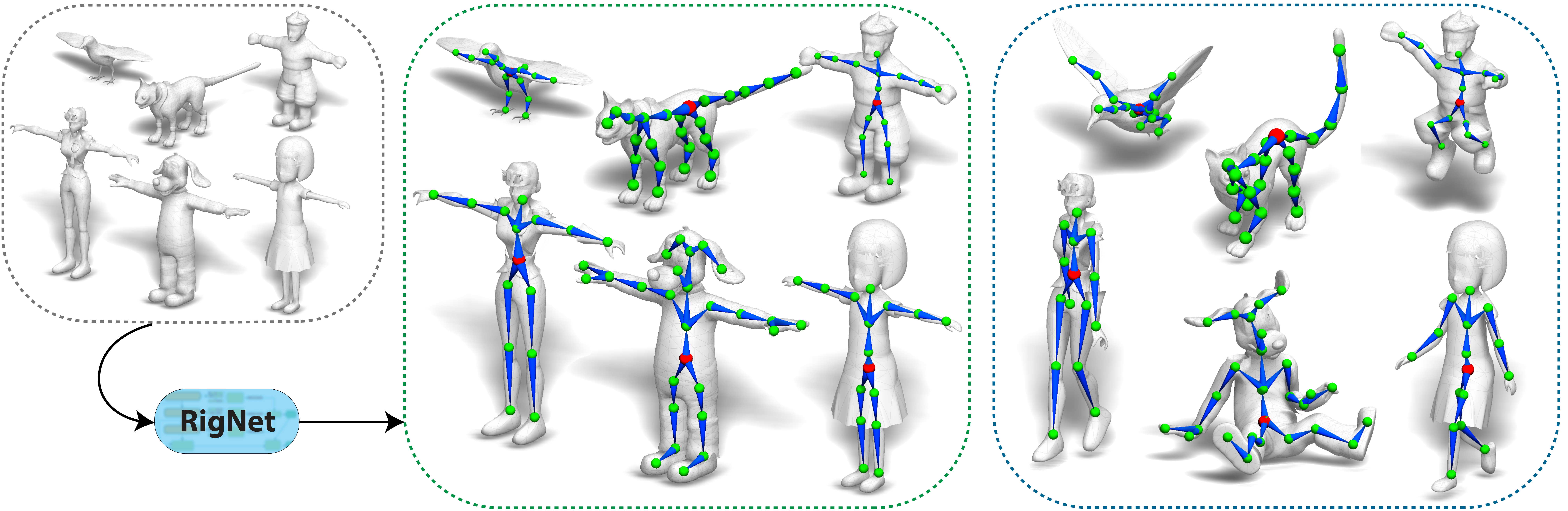}
\vskip -2mm
\caption{
Given a 3D character mesh, RigNet produces an animation skeleton and skin weights  tailored to the articulation structure of the input character. From left to right: input examples of test 3D meshes, predicted skeletons for each of them (joints  are shown in green and bones in blue),  and resulting skin deformations under different skeletal poses. Please see also our supplementary video: 
\textcolor{blue}{https://youtu.be/J90VETgWIDg}
}
\label{fig:teaser}
\end{teaserfigure}

\maketitle

\vspace{12mm}
\section{Introduction}
\input{chapters/introduction.tex}

\section{Related Work}
\input{chapters/related_work.tex}

\section{Overview}
\label{sec:overview}
\input{chapters/overview.tex}

\section{Method}
\label{sec:architecture}
\label{sec:method}
\input{chapters/architecture.tex}

\section{Training}
\label{sec:training}
\input{chapters/training.tex}

\section{Results}
\label{sec:results}
\input{chapters/results.tex}

\section{Limitations and Conclusion}
\input{chapters/conclusion.tex}

\begin{acks}
This research is partially funded by NSF (EAGER-1942069) and NSERC. Our experiments were performed in the UMass GPU cluster obtained under the Collaborative Fund managed by the Massachusetts Technology Collaborative. We thank Gopal Sharma, Difan Liu, and Olga Vesselova for their help and valuable suggestions. 
We also thank anonymous reviewers for their feedback.
\end{acks}

\bibliographystyle{ACM-Reference-Format}
\bibliography{references}

\vspace{5mm}
\appendix
\section{Appendix: Architecture details}
\input{chapters/suppl.tex}

\end{document}

%% file: symbols_format.tex
\newcommand{\ba}{\mathbf{a}}
\newcommand{\bb}{\mathbf{b}}

\newcommand{\bff}{\mathbf{f}}
\newcommand{\bg}{\mathbf{g}}
\newcommand{\bh}{\mathbf{h}}

\newcommand{\bm}{\mathbf{m}}

\newcommand{\bq}{\mathbf{q}}

\newcommand{\bs}{\mathbf{s}}
\newcommand{\bt}{\mathbf{t}}

\newcommand{\bv}{\mathbf{v}}
\newcommand{\bw}{\mathbf{w}}
\newcommand{\bx}{\mathbf{x}}

\newcommand{\bH}{\mathbf{H}}

\newcommand{\bS}{\mathbf{S}}

\newcommand{\bX}{\mathbf{X}}

\newcommand{\mM}{\mathcal{M}}

\newcommand{\mR}{\mathcal{R}}

\newcommand{\mN}{\mathcal{N}}

%% file: chapters/introduction.tex
 \label{sec:introduction}

There is a rapidly growing need for diverse, high-quality, animation-ready characters and avatars  in the areas of games, films, mixed Reality and social media. Hand-crafted character ``rigs'', where users create an animation ``skeleton'' and bind it to an input mesh (or ``skin''), have been the workhorse of articulated figure animation for over three decades. The skeleton represents the articulation structure of the character, and skeletal joint rotations provide an animator with direct hierarchical control of character pose.

We present a deep-learning based solution for automatic rig creation from an input 3D character. Our method predicts both a skeleton and skinning that match animator expectations (Figures~\ref{fig:teaser},~\ref{fig:gallery}).
In contrast to prior work that fits pre-defined skeletal templates of fixed joint count and topology to input 3D meshes  \cite{Baran:2007:ARA}, our method outputs skeletons more tailored to the underlying articulation structure of the input. Unlike pose estimation approaches designed for particular shape classes, such as humans or hands
\cite{Shotton2011,MoonCL18,HuangZLQX18,PavlakosZDD17a,HaquePLAYL16,XuGZC17}, 
our approach is not restricted by shape categorization or fixed skeleton structure. Our network represents a generic model of skeleton and skin prediction capable of rigging diverse characters (Figures \ref{fig:teaser},\ref{fig:gallery}).

\begin{figure}[t!]
  \centering
  \includegraphics[width=\linewidth]{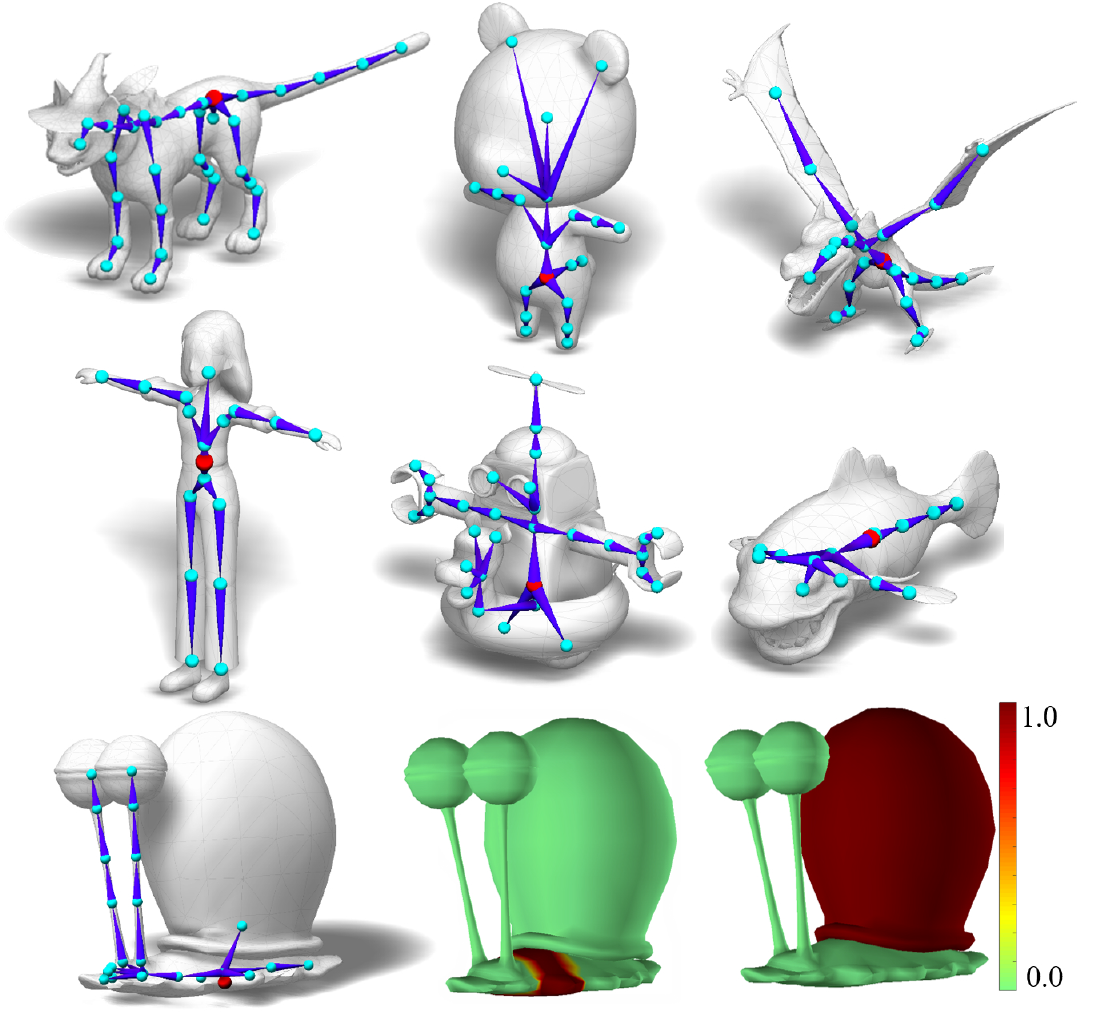}
  \vspace{-8mm}
  \caption{Examples of skeletons created by animators. In the bottom row, we show a rigged snail, including skinning weights for two of its parts.}
  \label{fig:training}
  \vspace{-4mm}  
\end{figure}

Predicting an animation skeleton and skinning from an arbitrary single static 3D mesh is an ambitious problem. As shown in Figure~\ref{fig:training},  animators create skeletons whose  number of joints and topology vary drastically across characters depending on their  underlying articulation structure. Animators also imbue an implicit understanding of creature anatomy into their skeletons. 
For example, character spines are often created closer to the back rather than the medial surface or centerline, mimicking human and animal anatomy  (Figure~\ref{fig:training}, cat); they will also likely introduce a proportionate elbow joint into cylindrical arm-like geometry  (Figure~\ref{fig:training}, teddy bear).
Similarly when computing skinning weights, animators  often perceive structures as highly rigid or smoother (Figure~\ref{fig:training}, snail).
An automatic rigging approach should ideally capture this animators' intuition about underlying moving parts and deformation. A\ learning approach is well suited for this task, especially if it is capable of learning from a large and diverse set of  rigged  models.

While animators largely agree on the skeletal topology and layout of joints for an input character, there is also some ambiguity both in terms of number and exact  joint placement (Figure~\ref{fig:variance}). For example, depending on animation intent, a hand may be represented using a single wrist joint or at a finer resolution with a hierarchy of hand joints (Figure~\ref{fig:variance}, top row). Spine and tail-like articulations may be captured using a variable number of joints (Figure~\ref{fig:variance}, bottom row). Thus, another challenge for a rigging method is to allow easy and direct control over the level-of-detail for the output skeleton.

To address the above challenges, we designed a  deep modular  architecture (Figure \ref{fig:architecture}). The first module is a graph neural network, trained to predict an appropriate number of joints and their placement, to capture the articulated mobility of the input character. As skeletal joint resolution can depend on the intended animation task, we provide users an optional parameter that can control the level-of-detail of the output skeleton (Figure~\ref{fig:control}).
A second module learns to predict a hierarchical tree structure (animation skeletons avoid cycles as a design choice) connecting the joints. The output bone structure is a function of   joints predicted from the first stage and shape features of the input character.
Subsequently, a third module, produces
a skinning weight vector per mesh vertex, indicating the degree of influence
it receives from different bones. This stage is also based on a graph neural network operating on shape features and  intrinsic distances from mesh vertices to the predicted bones.

 Our evaluation is three-fold: we show that \emph{RigNet} is  better than prior art when quantitatively compared to animator rigs (Tables~\ref{table:joint_comparison}, ~\ref{table:skinning_comparison}); qualitatively we show our rigs to be expressive and animation-ready (Figure~\ref{fig:teaser} and accompanying video); and technically, we evaluate the impact of various algorithm choices  on our output rigs
 (Tables~\ref{tab:joint_ablation},~\ref{tab:connectivity_ablition},~\ref{tab:skinning_ablation}). 

In summary, the contribution of this paper is an automated, end-to-end solution to the fundamentally important and challenging problem of  character rigging. Our technical contributions include a neural mesh attention and differentiable clustering scheme to localize joints, a  graph neural network for learning mesh representations, and a network that learns connectivity of graph nodes (in our case, skeleton joints). 
Our approach  significantly outperforms  purely geometric approaches \cite{Baran:2007:ARA}, and learning-based approaches that provide partial solutions to our problem i.e., perform only mesh skinning \cite{Liu2019}, or only skeleton prediction for volumetric inputs \cite{Xu19skeleton}.

%% file: chapters/related_work.tex
 In the following paragraphs, we discuss previous  approaches for producing animation skeletons, skin deformations  of 3D models, and graph neural networks.

\paragraph{Skeletons.}  Skeletal structures are fundamental representations in graphics and vision   \cite{Marr78,Dickinson:2009,Tagliasacchi16}. Shape skeletons vary in concept from precise geometric constructs like the  medial axis representations \cite{Blum,Amenta:1998:SRV,Attali:1997:CSC,Siddiqi08}, curvilinear representations or meso-skeletons
 \cite{Singh1998WiresAG,Au:2008:SEM,Cao10,Tagliasacchi:2009:CSE,Huang:2013:LMS,yin2018p2pnet}, to piecewise linear structures \cite{Katz2003,Song96,Siddiqi:1999:SGS,Hilaga01}. Our work is mostly related to  animator-centric skeletons \cite{Magnenat-Thalmann:1989}, which are  designed to capture the mobility of an articulated shape. As discussed in the previous section, apart from shape geometry, the placement of joints and bones in animation skeletons is driven by the animator's understanding of character's anatomy and  expected deformations.

The earliest approach to automatic rigging  of input 3D models is the pioneering method of ``Pinocchio'' \cite{Baran:2007:ARA}. Pinocchio follows a combination of discrete and continuous optimization to fit a pre-defined skeleton template to a 3D model, and also performs skinning through heat diffusion.  Fitting tends to fail when the input shape structure is incompatible with the selected template. Hand-crafting templates for every possible structural variation of an input character is cumbersome. More recently, inspired by 3D\ pose estimation approaches \cite{HaquePLAYL16,PavlakosZDD17a,Newell2016StackedHN,ge2018_Point,MoonCL18,HuangZLQX18,Wan_2018_CVPR}, Xu et al. \cite{Xu19skeleton} proposed learning a volumetric network for producing skeletons, without skinning, from input 3D characters. Pre-processing the input mesh to a coarser voxel representation can: eliminate surface features (like elbow or knee protrusions) useful for accurate joint detection and placement; alter the input shape topology (like proximal fingers represented as a voxel mitten); or accumulate approximation errors.
{\it RigNet} compares favorably to these methods (Figure~\ref{fig:comparison_skeleton}, Table~\ref{table:joint_comparison}), without requiring pre-defined skeletal templates, pre-processing or lossy conversion between shape representations.  

\paragraph{Skin deformations.} A wide range of approaches have also been proposed to model skin deformations, ranging from 
 physics-based methods \cite{Kim17datadriven,Mukai16,Si15,Komaritzan2018,Komaritzan2019FastPS},
geometric methods \cite{Kavan05,Kavan07,Kavan12,Wareham08,Jacobson11,Bang18,Dionne13,Dionne14},
to data-driven methods that produce skinning from a sequence of examples \cite{Loper15,Le14seq,Doug05,qiao2018learning}. Given a single input character, it is  common to resort to geometric methods for skin deformation, such as Linear Blend Skinning (LBS) or Dual Quaternion Skinning (DQS) \cite{Kavan07,Binh16} due to their simplicity and computational efficiency. These methods require input skinning weights per vertex which are either interactively painted and edited \cite{Bang18}, or automatically estimated based on hand-engineered functions of shape geometry and skeleton \cite{Baran:2007:ARA,Kavan12,Wareham08,Jacobson11,Bang18,Dionne13,Dionne14}. It is difficult for such geometric approaches to account for any anatomic considerations implicit in input meshes, such as the disparity between animator and geometric spines, or the skin flexibility/rigidity of different articulations.

Data-driven approaches like ours, however, can capture  anatomic insights present in animator-created rigs. Neuroskinning \cite{Liu2019} attempts to learn skinning from an input family of 3D characters. Their network performs graph convolution by learning edge weights within mesh neighborhoods, and outputting vertex features as weighted combinations of neighboring vertex features. Our method instead learns edge feature representations within both mesh and geodesic neighborhoods, and combines them into vertex representations inspired by the edge convolution scheme of \cite{dgcnn}. Our network input uses intrinsic shape representations capturing geodesic distances between vertices and bones, rather than relying on extrinsic features, such as Euclidean distance. Unlike Neuroskinning, our method  does not require any input joint categorization during training or testing.
Most importantly, our method proposes a complete solution (skeleton and skinning) with better results (Tables~\ref{table:joint_comparison}, ~\ref{table:skinning_comparison}).

We note that our method is complementary to  physics-based or deep learning methods that produce non-linear deformations, such as muscle bulges, on top of skin deformations \cite{Mukai16,Bailey18,Luo2018DeepWarpDN}, or  rely on input bones and skinning weights to compute other deformation approximations
\cite{jeruzalski2019nasa}.
These methods require input bones and skinning weights that are readily provided by our method.  

\begin{figure}[t!]
  \centering
  \includegraphics[width=\linewidth]{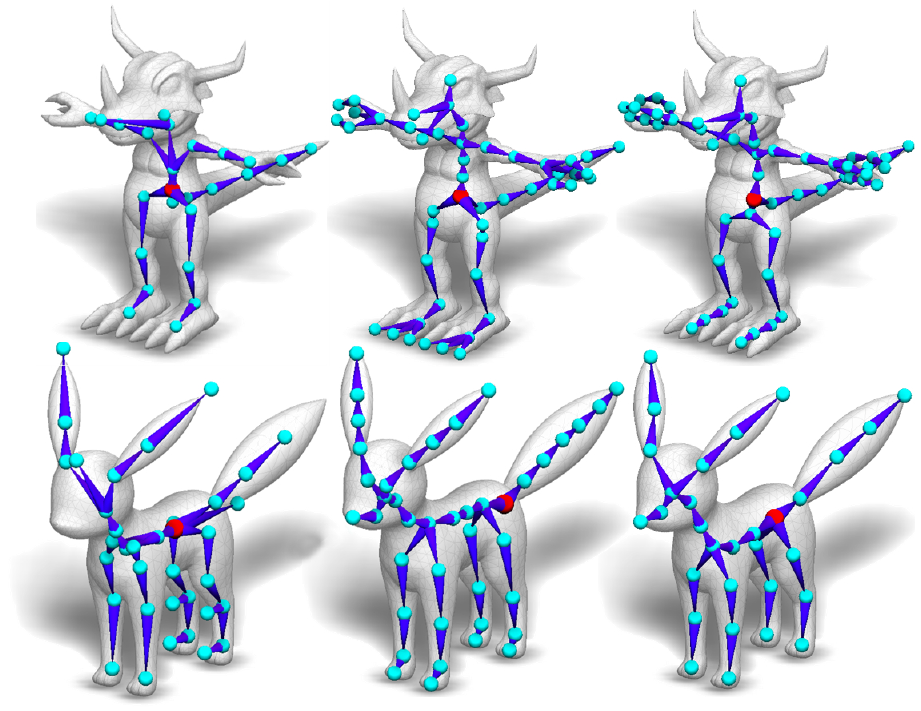}
  \vspace{-6mm}
  \caption{Models rigged by three different artists. Although they tend to agree on  skeleton layout and expected articulation, there is variance in terms of number of joints and overall level-of-detail.}
  \label{fig:variance}
  \vspace{-4mm}  
\end{figure}

\paragraph{Graph Neural Networks.} Graph Neural Networks (GNNs) have become increasingly popular for graph
processing tasks  \cite{wu2019comprehensive,Scarselli:2009:GRP,Bruna2013spectral,Henaff2015spectral,kipf2016semisupervised,Defferrard2016spectral,Li:2015:Gated,Battaglia:2016:PYS,hamilton2017inductive,Hamilton:2017:survey}. Recently, GNNs have also been proposed for geometric deep learning 
on point sets \cite{dgcnn},  
meshes \cite{masci2015geodesic,hanocka2019meshcnn}, intrinsic or spectral representations 
\cite{bronstein2017geometric,Boscaini2016,Monti2017,yi2017syncspeccnn}. Our graph neural network adapts the operator proposed in \cite{dgcnn} to perform edge convolutions within mesh-based and geodesic neighborhoods. Our network also weighs and combines representations from  mesh topology, local and global shape geometry.
Notably, our approach judiciously combines several other neural modules for detecting and connecting joints, with a graph neural network, to provide an integrated deep  architecture for end-to-end character rigging.

%% file: chapters/overview.tex
\begin{figure*}[t!]
  \centering
  \includegraphics[width=\linewidth]{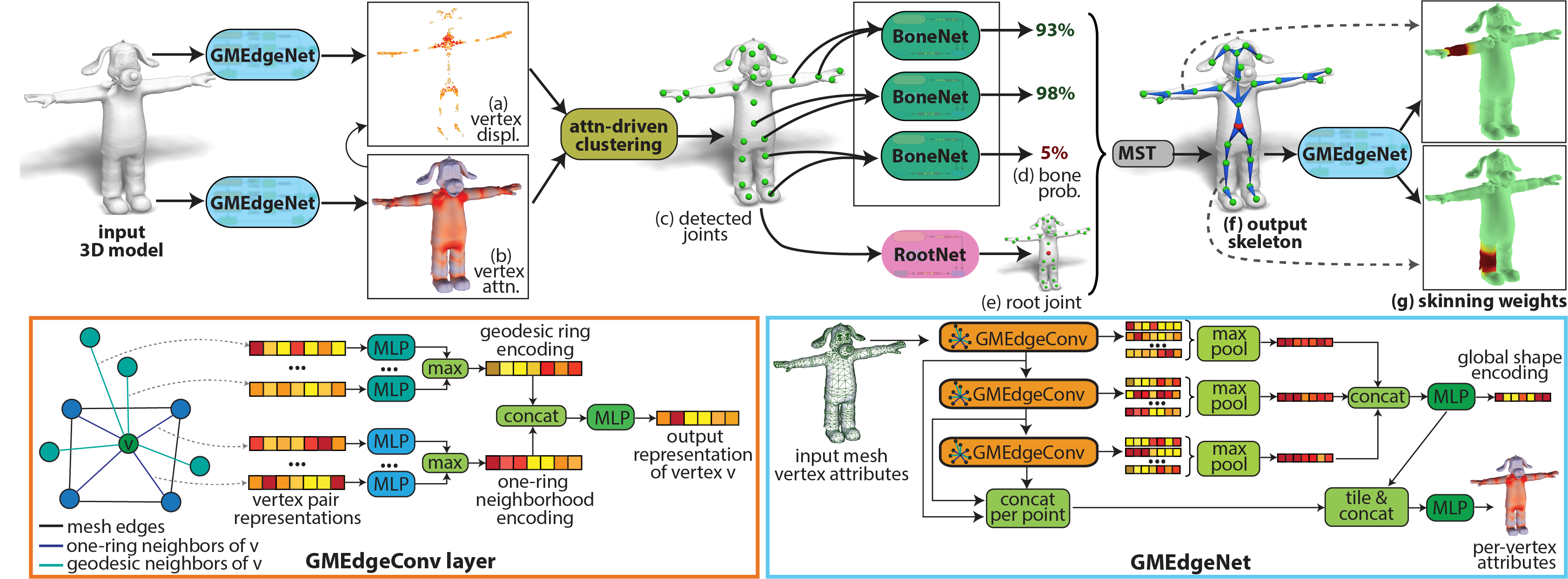}
  \vspace{-7mm}  
  \caption{Top: Pipeline of our method. (a) Given an input 3D model, a graph neural network, namely GMEdgeNet,  predicts displacements of vertices towards neighboring joints. (b) Another GMEdgeNet module with separate parameters predicts an attention function over the mesh  that indicates areas  more relevant for joint prediction (redder values indicate stronger attention - the displaced vertices are also colored according to attention). (c) Driven by the mesh attention, a clustering module detects joints shown as green balls. (d) Given the detected joints, a neural module (BoneNet, see also Figure \ref{fig:bonenet}) predicts probabilities for each pair of joints to be connected. (e) Another module (RootNet) extracts the root joint. (f) A Minimum Spanning Tree (MST) algorithm uses the BoneNet and RootNet outputs to form an animation skeleton. (g) Finally, a GMEdgeNet module  outputs the skinning weights based on the predicted skeleton. Bottom: Architecture of GMEdgeNet, and its graph convolution layer (GMEdgeConv). 
  }  
\label{fig:architecture}
\vspace{-3mm}  
\end{figure*}

Given an input 3D mesh of a character, our method predicts an animation skeleton and skinning tailored for its underlying articulation structure and geometry.  Both the skeleton and skinning weights are animator-editable primitives that can be further refined through standard modeling and animation pipelines. Our method is based on a deep  architecture (Figure \ref{fig:architecture}), which operates directly on the mesh representation. We do not assume known input character class, part structure, or skeletal joint categories during training or testing.
 Our  only assumption is that the input training and test shapes have a consistent upright and frontfacing orientation. Below, we briefly overview the key
aspects of our architecture. In Section \ref{sec:method}, we explain its stages  in more detail. 

\paragraph{Skeletal joint prediction.} The first module of our architecture is trained to predict the location of joints that will be used to form the animation skeleton.
To this end, it learns to displace mesh geometry towards candidate joint locations (Figure \ref{fig:architecture}a).
The module is based on a graph neural network, which extracts topology- and geometry-aware
features from the mesh to learn these displacements. A\ key idea of our architecture  in this stage is to learn a weight function over the input mesh, a form of neural mesh attention, which is used to reveal which surface areas are more relevant for localizing joints (Figure \ref{fig:architecture}b). Our experiments demonstrate that this leads to more accurate skeletons. The displaced mesh geometry tends to form  clusters around candidate joint locations. We introduce a differentiable clustering scheme, which uses the neural mesh attention,  to extract the joint locations  (Figure \ref{fig:architecture}c). 

Since  the final animation skeleton may depend on the task or the artists' preferences, our method also allows  optional user input in the form of a single parameter to control the level-of-detail, or granularity, of the output skeleton. For example, some applications, like crowd simulation, may not require rigging of small parts (e.g., hands or fingers), while other applications, like FPS games, rigging such parts is more important. By controlling a single parameter through a slider, fewer or more joints are introduced to
capture different level-of-detail for the output skeleton
(see Figure \ref{fig:control}).

\vspace{-1mm}
\paragraph{Skeleton connectivity prediction.} The next module in our architecture learns which pairs of extracted joints should be connected with bones. Our module takes as input  the predicted joints from the previous step, including a learned shape and skeleton representation, and outputs a probability representing whether each pair should be connected with a bone or not (Figure \ref{fig:architecture}d). We found that learned joint and shape  representations are important to reliably estimate bones, since the skeleton connectivity depends not only on joint locations but also the overall shape and skeleton geometry. The  bone probabilities are used as input to a Minimum Spanning Tree algorithm that prioritizes the most likely bones to form a  tree-structured skeleton, starting from a root joint picked from another trained neural module (Figure \ref{fig:architecture}e).

\vspace{-1mm}
\paragraph{Skinning prediction.} Given a predicted skeleton (Figure \ref{fig:architecture}f),
the last module of our architecture produces a weight vector per mesh vertex indicating the degree of influence it receives from different bones (Figure \ref{fig:architecture}g). Our method  is inspired by Neuroskinning \cite{Liu2019}, yet, with important differences in the architecture, bone and shape representations, and the use of volumetric geodesic distances from vertices to bones (as opposed to Euclidean distances). 

\vspace{-1mm}
\paragraph{Training and generalization.} Our architecture is trained via a combination of loss functions measuring deviation in joint locations, bone connectivity, and skinning weight differences with respect to the training skeletons. Our architecture is trained on input characters that vary significantly in terms of structure, number and geometry of moving parts e.g., humanoids, bipeds, quadrupeds, fish, toys, fictional characters. Our test set is also similarly diverse. We observe that our method is  able to generalize to characters with different number of underlying articulating parts (Figure \ref{fig:gallery}).

%% file: chapters/architecture.tex
We now explain our architecture (Figure \ref{fig:architecture}) for rigging an input 3D model at test time in detail.
In the following subsections, we discuss each stage of our architecture.  Then in Section \ref{sec:training}, we discuss training. 

\subsection{Joint prediction} 
\label{sec:joints}
Given an input mesh $\mM$, the first stage of our architecture outputs a set of 3D\ joint locations $\bt=\{ \bt_i \}$, where $t_i \in \mR^3$. 
One particular complication related to this mapping is that the number of articulating parts, and in turn, the number of joints is not the same for all characters. For example, a multiped creature is expected to have more joints than a biped. We use a combination of regression and adaptive clustering to solve for the joint locations and their number. In the regression step, the mesh vertices are displaced towards their nearest candidate joint locations. This step results in accumulating points near joint locations (Figure \ref{fig:architecture}a). The second step localizes the joints by clustering the displaced points and setting the cluster centers as joint locations
(Figure \ref{fig:architecture}b).
The number of resulting clusters is determined adaptively according to the underlying point density and learned clustering parameters. Performing clustering without first displacing the vertices  fails to extract reasonable joints, since the original position of mesh vertices is often far from joint locations.
In the next paragraphs, we explain the regression and clustering steps.
\vspace{-1mm}
\paragraph{Regression.} In this step, the mesh vertices are regressed to their nearest candidate joint locations. This is performed through a learned neural network function that takes  as input the mesh $\mM$ and outputs \emph{vertex displacements}. Specifically, given the original mesh vertex locations $\bv$, our   displacement module $f_d$ outputs perturbed points $\bq$:
\begin{equation}
\bq = \bv +\ f_d( \mM; \bw_d)
\end{equation}
where $\bw_d$ are  learned parameters of this module. Figure \ref{fig:architecture}a visualizes displaced points for a characteristic example. This mapping is  reminiscent of  \mbox{P2P-Net} \cite{yin2018p2pnet} that learns to displace surface points across different domains e.g., surface points to meso-skeletons.\ In our case, the goal is to map mesh vertices to joint locations.  An important aspect of our setting is that not all surface points are equally useful to determine joint locations e.g., the  vertices located near the elbow region of an arm are more likely to reveal  elbow joints compared to other vertices.  Thus, we also designed a  neural network function $f_a$ that outputs an attention map which represents a confidence of localizing a joint from each vertex. Specifically, the attention map $\ba=\{a_v\}$ includes a scalar value per vertex, where $a_v \in [0,1]$, and is computed as follows:
\begin{equation}
\ba = \ f_a( \mM; \bw_a)
\end{equation}
where $\bw_a$ are learned parameters of the attention module.  Figure \ref{fig:architecture}b visualizes the map for a characteristic example.

\begin{figure}[t!]
  \centering
  \includegraphics[width=\linewidth]{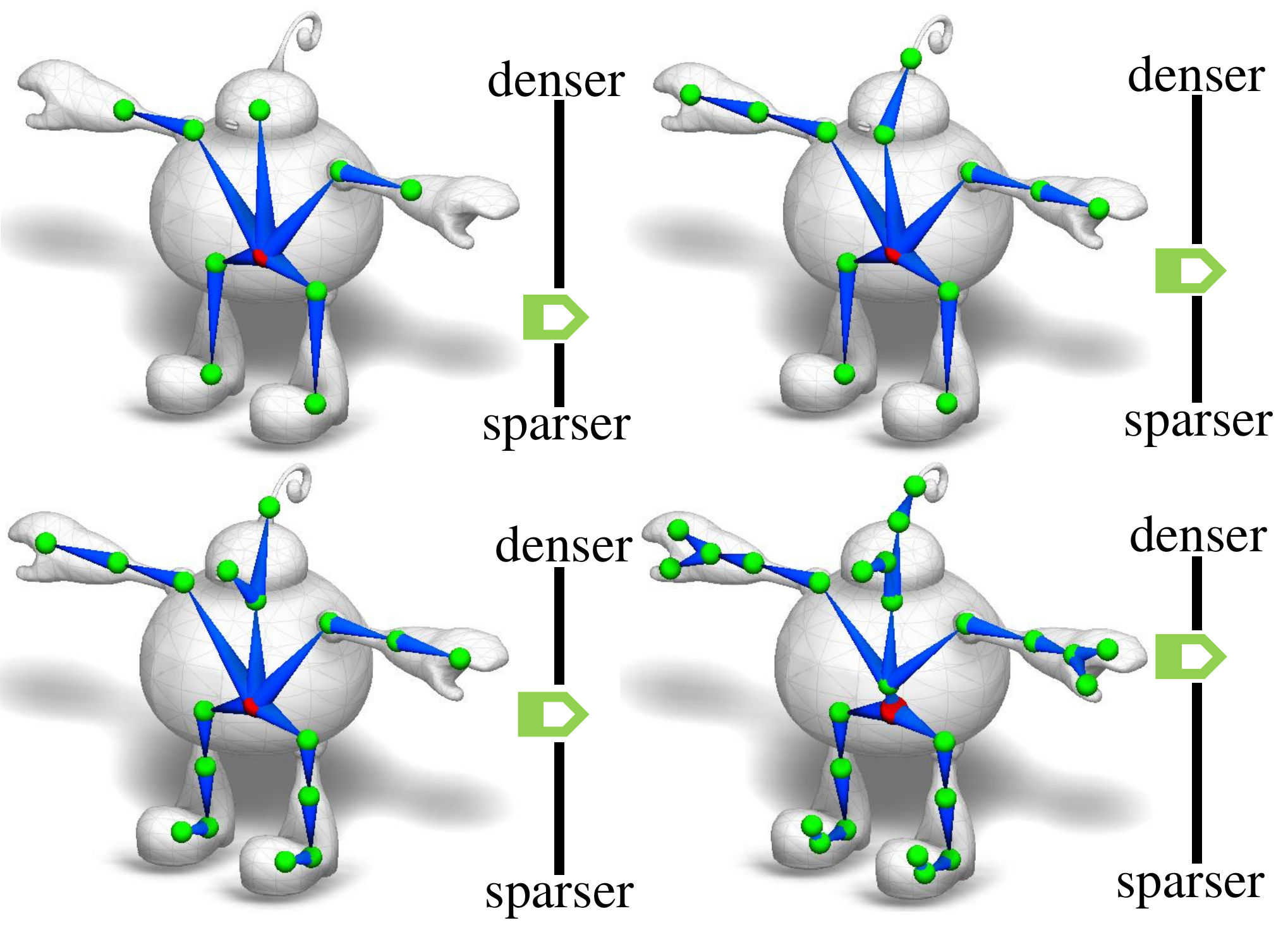}
  \vspace{-7mm}
  \caption{Effect of increasing the bandwidth parameter that controls the level-of-detail, or granularity, of our predicted skeleton.}
  \label{fig:control}
  \vspace{-3mm}  
\end{figure}

\vspace{-1mm}
\paragraph{ Module internals.} Both displacement and attention neural network modules operate  on the mesh graph.  As we show in our experiments, operating on the mesh graph yields significantly better performance
compared to using alternative architectures that operate on point-sampled representations \cite{yin2018p2pnet}  or volumetric representations  \cite{Xu19skeleton}.
Our networks builds upon the edge convolution  proposed in \cite{dgcnn}, also known  as `EdgeConv''. Given feature vectors $\bX=\{\bx_v\}$ at mesh vertices, the output of an EdgeConv
operation at a vertex  is a new feature vector encoding its local graph neighborhood: 
\mbox{$\bx'_v = \max \limits_{u \in \mN(v)} MLP(\bx_v, \bx_u - \bx_v; \bw_{mlp})$} where $MLP$ denotes a learned multi-layer perceptron, $\bw_{mlp}$ are  its learned parameters, and $\mN(v)$ is the graph neighborhood of vertex $v$. Defining a proper graph neighborhood for our task turned out to be fruitful. One possibility is to simply use one-ring vertex neighborhoods for edge convolution. We instead found that this strategy makes the network  sensitive to the input mesh tessellation and results in lower performance. Instead, we found that it is better to
define the graph neighborhood of a vertex by considering both its one-ring mesh neighbors, and also  the vertices located within a geodesic ball centered  at it. We also found that it is better to learn separate MLPs for
 mesh and geodesic neighborhoods, then concatenate their outputs and process them through another MLP. In this manner, the networks
learn to weigh the importance of  topology-aware features over more geometry-aware ones. Specifically, our convolution operator, called GMEdgeConv (see also Figure \ref{fig:architecture}, bottom)
  is defined as follows:
\begin{eqnarray}
\bx_{v,m} = \max \limits_{u \in \mN_m(v)} MLP(\bx_v, \bx_u - \bx_v; \bw_m) \\
\bx_{v,g} = \max \limits_{u \in \mN_g(v)} MLP(\bx_v, \bx_u - \bx_v; \bw_g) \\
\bx_v' =MLP( concat( \bx_{v,m}, \,\, \bx_{v,g} ); \bw_c )
\end{eqnarray}
where $\mN_m(v)$ are the one-ring mesh neighborhoods of vertex $v$, $\mN_g(v)$ are the  vertices  from its geodesic ball. In all our experiments, we used a ball radius $r=0.06$ of the longest dimension of the model, which is tuned through grid search in a hold-out validation set. The weights $\bw_m$, $\bw_g$, and  $\bw_c$ are learned parameters for the above MLPs. We note that we experimented with the attention mechanism proposed in \cite{Liu2019}, yet we did not find any significant improvements. This is potentially due to the fact that EdgeConv already learns edge representations based on the pairwise functions of vertex features, which may implicitly encode edge importance.

Both the vertex displacement and attention modules start with the vertex positions as input features.
They share
the same internal architecture, which we call GMEdgeNet   (see also Figure \ref{fig:architecture}, bottom). GMEdgeNet stacks three GMEdgeConv layers,  each followed with a global max-pooling layer.
The representations from each pooling layer are concatenated to form a global mesh representation. The output per-vertex representations from all GMEdgeConv layers, as well as the global mesh representation, are further concatenated, then processed  through a 3-layer MLP. In this manner, the learned vertex representations incorporate both local and global information. In the case of the vertex displacement module, the feature representation  are transformed to 3D displacements per each vertex through another MLP. In the case of the vertex attention module, the per-vertex feature representations  are transformed through a MLP and a sigmoid non-linearity to produce a scalar attention value per vertex. Both modules use their own set of learned parameters for their GMEdgeConv layers and  MLPs. More details about their architecture are provided in the appendix.

\vspace{-1mm}
\paragraph{Clustering.} This step takes as input the displaced points $\bq$ along with their corresponding attention values  $\ba$, and outputs joints. As shown in Figure \ref{fig:architecture}a,  points
tend to concentrate in areas around candidate joint locations. Areas with higher point density and greater  attention
are strong  indicators of joint presence. We resort to density-based clustering to detect local maxima of point density and use those as joint locations. In particular, we employ a variant of mean-shift clustering, which also uses our learned attention map. A particular advantage of mean-shift clustering is that it does not explicitly require as input the number of target clusters.

 In classical mean-shift clustering \cite{Cheng95}, each data point is equipped with a kernel function. The sum of  kernel functions results in a continuous density estimate, and the local maxima (modes) correspond to cluster centers. Mean-shift clustering is performed iteratively; at each iteration, all points are shifted towards  density modes.
In our implementation, the kernel  is also modulated by the vertex attention. In this manner, points with greater attention influence the estimation of density  more. Specifically, at each mean-shift iteration, each points is displaced according to the vector:
\begin{equation}
\bm_v = \frac{ \sum\limits_{u} a_{u} \cdot  K(\bq_u - \bq_v, h) \cdot \bq_u  } {\sum\limits_{u} a_{u} \cdot  K(\bq_u - \bq_v, h)   }  - \bq_v
\label{eq:mean-shift}
\end{equation}
where $K(\bq_u - \bq_v, h)=max(1 - ||\bq_u - \bq_v||^2/{h^2}, 0)$ is the Epanechnikov kernel with learned bandwidth $h$. We found that the Epanechnikov kernel produces better  clustering results than a Gaussian kernel or a triangular kernel. The  mean-shift iterations are implemented through a recurrent module in our  architecture, similarly to the recurrent pixel grouping  in Kong and Fowlkes \shortcite{kong2018grouppixels}, which   also enables training of the bandwidth through backpropagation.

At test time, we perform mean-shift iterations until convergence (i.e., no point is shifted for a Euclidean distance more than $10^{-3}$). As a result, the shifted points ``collapse'' into distinct modes (Figure \ref{fig:architecture}c). To extract these modes, we start with the point with highest density, and remove  all its neighbors within radius equal to the bandwidth $h$.  This point represents a  mode, and we create a joint at its location. Then we proceed by finding the point with the second largest density among the remaining ones, suppress its neighbors, and create another joint. This process continues until no other points remain. The output of the step are the modes that correspond to the the set of detected joints $\bt=\{ \bt_i \}$. 

\vspace{-1mm}
\paragraph{User control.}  Since animators may prefer to have  more  control over the placement of joints, we allow them to override the learned bandwidth value, by interactively manipulating a slider controlling its value (Figure \ref{fig:control}). We found that modifying the bandwidth directly affects the level-of-detail of the output skeleton.
Lowering the bandwidth results in denser joint placement, while increasing it results in sparser skeletons. We note that the bandwidth cannot be set to arbitrary values e.g., a zero bandwidth value will cause each displaced vertex to become a joint. In our implementation, we empirically set an editable range from 0.01 to 0.1.
The resulting joints can be processed by the next modules of our architecture to produce the bone connectivity and skinning based on their updated positions.

\vspace{-1mm}
\paragraph{Symmetrization.} 3D characters are often modeled based on a neutral pose (e.g., ``T-pose''), and as a result their body shapes usually have bilateral symmetry. In such cases, we symmetrize joint prediction by reflecting the displaced points $\bq$ and attention map $\ba$
 according to the global bilateral symmetry plane  before performing clustering. As a result, the joint prediction is more robust to any small  inconsistencies produced in either side.

\subsection{Connectivity prediction} 
\begin{figure}[t!]
  \centering
  \includegraphics[width=\linewidth]{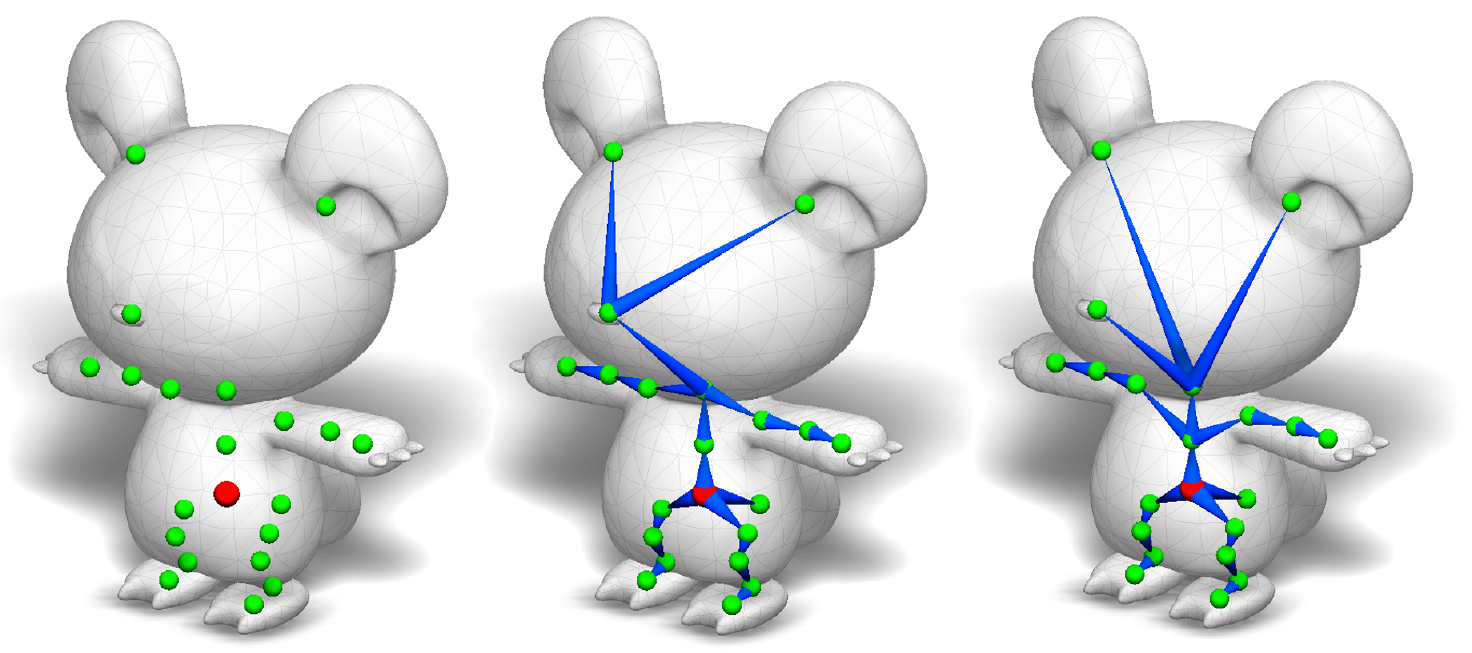}
\vspace{-7mm}  
  \caption{ \emph{Left:} Joints detected by our method.  The root joint is shown in red. \emph{Middle:} Skeleton created with Prim's algorithm based on Euclidean distances as edge cost. \emph{Right:} Skeleton created using the negative log of BoneNet probabilities as cost.}
\vspace{-3mm}
  \label{fig:mst}
\end{figure}

Given the joints extracted from the previous stage, the  connectivity prediction stage determines how these joints should be connected to form the animation skeleton. At the heart of this stage lies
a  learned neural module that outputs the probability of connecting each pair of joints via a bone. These pairwise bone probabilities are used as input to Prim's algorithm that creates a Minimum Spanning Tree (MST) representing the animation skeleton. We found that using these bone probabilities to extract the MST\ resulted in   skeletons that  agree  with animator-created ones  more in topology compared to simpler schemes e.g., using Euclidean distances between joints (see Figure \ref{fig:mst} and experiments). In the following paragraphs, we explain the module for determining the bone probabilities for each pair of joints, then we discuss the cost function used for creating the MST. 

\setlength{\columnsep}{10pt}
\begin{wrapfigure}{R}{0.55\linewidth}
 \vspace{-2mm} 
  \includegraphics[width=1\linewidth]{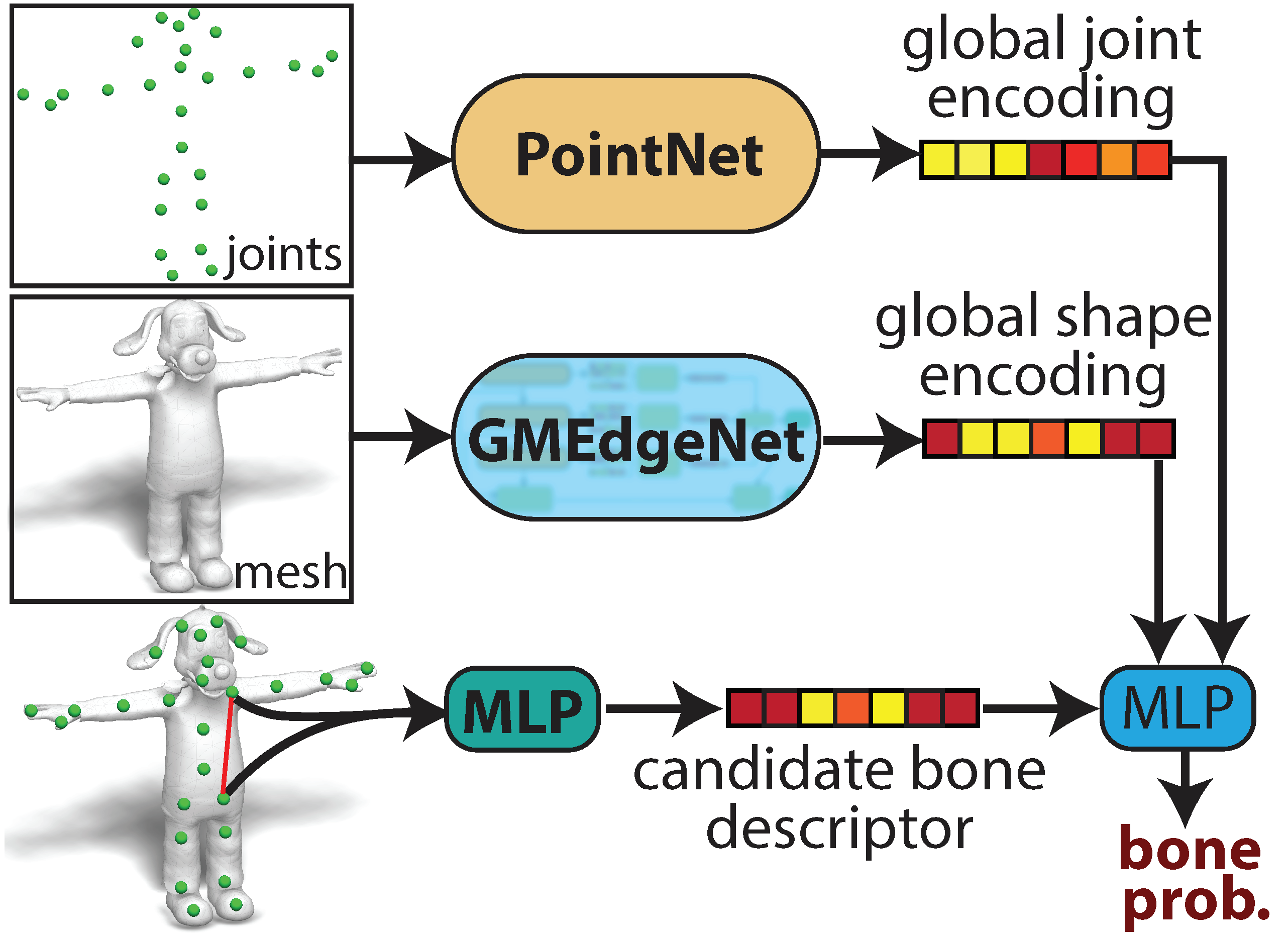}
  \vspace{-8mm}   
  \caption{BoneNet architecture.}
\vspace{-4mm}     
  \label{fig:bonenet}
\end{wrapfigure}

\vspace{-1mm}
\paragraph{Bone module.} The bone module, which we call ``BoneNet'', takes as input our predicted joints $\bt$ along with the input mesh $\mM$, and outputs the probability $p_{i,j}$ for connecting each pair of joints via a bone. By processing all pairs of joints through the same module, we extract a pairwise matrix representing all candidate bone probabilities. The architecture of the  module is shown in Figure \ref{fig:bonenet}. For each pair of joints, the module processes three representations that capture global shape geometry, skeleton geometry, and features from the  input pair of joints. In our experiments, we found that this combination offered the best bone prediction performance. More specifically, BoneNet takes as input:  (a) a $128$-dimensional representation $\bg_s$ encoding global shape geometry, which is extracted from the max-pooling layers of GMEdgeNet (see also Figure \ref{fig:architecture}, bottom),
  (b) a $128$-dimensional representation $\bg_t$ encoding the overall skeleton geometry by treating joints  as a collection of points and using  a learned PointNet to produce it \cite{qi2017pointnetplusplus}, and  (c) a representation encoding  the input pair of joints. To produce this last representation, we first concatenate the positions of two joints $\{\bt_i, \bt_j\}$,  their Euclidean distance $d_{i,j}$, and another scalar $o_{i,j}$ capturing the proportion of the candidate bone lying in the exterior of the mesh. The Euclidean distance and proportion are useful  indicators of joint connectivity:  the smaller the distance between two joints,
the
more likely is a bone between them. If the candidate bone protrudes significantly outside the shape, then it is less likely to choose it for the final skeleton.  We transform the raw features $[\bt_i, \bt_j, d_{i,j}, o_{i,j}]$ into a $256$-dimensional \textit{bone representation $\bff_{i,j}$} through a MLP. The bone probability is computed via a 2-layer MLP  operating on the concatenation of these three representations, followed by a sigmoid: 
\begin{equation}
p_{i,j}  = sigmoid \big( MLP( \bff_{i,j}, \bg_s, \bg_t; \bw_b)  \big)
\label{eq:bonenet}
\end{equation}
where $\bw_b$ are learned module parameters.  Details about the architecture of  BoneNet are provided in the appendix. 

\vspace{-1mm}
\paragraph{Skeleton extraction.} The skeleton extraction step aims to infer the most likely tree-structured animation skeleton among all possible candidates. If we consider the choice of selecting an edge  in a tree as an independent random variable, the joint probability of a tree is equal to the product of its edge probabilities. Maximizing the joint probability is equivalent to minimizing the negative log probabilities of the edges: $w_{i,j} = -\log p_{i,j}$. Thus, by defining a dense graph whose nodes are the extracted joints, and edges have weights  $w_{i,j}$, we can use a MST algorithm to solve this problem. In our implementation, we use Prim's algorithm
\cite{Prim1957}. Any joint can serve as a starting, or root joint for Prim's algorithm. However, since the root joint is used to control the global  character's body position and orientation and is important for motion re-targeting tasks, this stage also predicts which joint should be used as root. One common choice is to select the joint closer to the center of gravity for the character. However, we found that this choice is not always consistent with animators' preferences (Figure \ref{fig:training}, root nodes in the cat and dragon are further away from their centroids). Instead, we found that the selection of the root joint can also be performed more reliably using a  neural module. Specifically, our method incorporates a module, which was call RootNet. Its internal architecture follows BoneNet. It takes  as input the global shape representation $\bg_s$ and global joint representation $\bg_t$ (as in BoneNet). It also takes as input a  joint representation  $\bff_i$ learned through a MLP operating on its location and distance  $d_{i,c}$ to the bilateral symmetry plane. The latter feature was driven by the observation that root joints are often placed along this symmetry plane.
RootNet
outputs the root joint probability as follows:
\begin{equation}
p_{i,r} = softmax \big( MLP( \bff_i, \bg_s, \bg_t; \bw_r)  \big)
\label{eq:rootnet}
\end{equation} 
where $\bw_r$ are learned parameters. At test time, we select the joint with highest probability as root joint to initiate the Prim's algorithm.

\subsection{Skinning prediction} 

After producing the animation skeleton, the final stage of our architecture is the prediction of skinning weights for each mesh vertex to complete the rigging process. To perform skinning, we first extract a  mesh representation  capturing the spatial relationship  of mesh vertices with respect to the skeleton. The representation is inspired by previous skinning methods \cite{Dionne13, Jacobson11} that compute influences of bones on vertices according to volumetric geodesic distances between them. This mesh representation is processed through a graph neural network that outputs the per-vertex skinning weights. In the next paragraphs, we describe the representation and  network.

\vspace{-1mm}
\paragraph{Skeleton-aware mesh representation.} The first step of the skinning stage is to compute a mesh representation $\bH=\{\bh_v\}$, which stores a feature vector for each mesh vertex $v$ and captures its spatial relationship with respect to the skeleton. Specifically, for each vertex we compute volumetric geodesic distances to all the bones i.e, shortest path lengths from vertex to bones passing through the interior mesh volume. We use a  implementation that approximates the volumetric geodesic distances based on \cite{Dionne13}; other potentially more accurate approximations could also be used \cite{Crane13,Solomon14}. Then for each vertex $v$, we sort the bones according to their volumetric geodesic distance to it, and create an ordered feature sequence $\{\bb_{r,v}\}_{r=1...K}$, where $r$ denotes an index to the sorted list of bones. Each feature vector $\bb_{r,v}$  concatenates the 3D positions of the starting and end joints of bone $r$, and the inverse of the volumetric geodesic distance from the vertex $v$ to this bone ($1/{D_{r,v}}$). The reason for ordering the bones wrt each vertex is to promote consistency in the resulting representation i.e., the first entry represents always the closest bone to the vertex, the second entry represents the second closest bone, and so on. 
In our implementation, we use the $K=5$ closest bones selected based on hold-out validation. If a skeleton contains less than $K$ bones, we simply repeat the last bone in the sequence. The final per-vertex representation $\bh_v$ is formed by concatenating the vertex position and above ordered sequence $\{\bb_{r,v}\}_{r=1...K}$. 

\vspace{-1mm}
\paragraph{Skinning module} The  module $f_s$  transforms the above skeleton-aware mesh representation $\bH$ to  skinning weights $\bS=\{\bs_v\}$: 
\begin{equation}
\bS = \ f_s( \bH; \bw_s)
\end{equation}
where $\bw_s$ are learned parameters. The skinning network follows GMEdgeNet. The last layer outputs a $1280$-dimensional per-vertex feature vector, which is transformed to a per-vertex skinning weight vector $\bs_v$ through a learned MLP and a softmax function. This ensures that the skinning weights for each vertex are positive and sum to $1$. The entries of the output skinning weight vector $\bs_v$ are ordered according to the volumetric geodesic distance of the vertex $v$ to the corresponding bones.

%% file: chapters/training.tex
The goal of our training procedure is to learn the parameters of the networks used in each of the three stages of \emph{RigNet}. Training is performed on a dataset of rigged characters described in Section \ref{sec:results}. %We first train each of the stages of the architecture individually, then we fine-tune them jointly. 

\subsection{Joint prediction stage training} 
Given a set of 
training characters, each with
skeletal joints $\hat{\bt}=\{ \hat{\bt}_k \}$, we learn the parameters $\bw_a$, $\bw_d$, and bandwidth $h$ of this stage such that the estimated skeletal joints approach as closely as possible to the training ones. Since the estimated skeletal joints originate from  mesh vertices that  collapse into modes after mean shift clustering, we can alternatively formulate the above learning goal as a problem of minimizing the distance of collapsed vertices
to nearest 
 training joints and vice versa. Specifically, we minimize   the symmetric Chamfer distance between collapsed vertices $\{\bt_v\}$ and training joints $\{\hat{\bt}_k\}$:
\begin{equation}
L_{cd}(\bw_a, \bw_d, h)  =  \frac{1}{V}  \sum_{v=1}^V  \min\limits_{k} || \bt_v - \hat{\bt}_k ||
 +   
\frac{1}{K}  \sum_{k=1}^K \min\limits_{v} || \bt_v - \hat{\bt}_k ||
\label{eq:cluster_sup}
\end{equation}
%We note that we did not divide the sums with the number of ground-truth or estimated joints. This implicitly adds more loss for skeletons with more joints, putting more emphasis on training characters with more detailed skeletons. The loss also tends to increase when the number of estimated joints becomes significantly higher than the number of training ones due to joint misplacements, which is also helpful for learning a reasonable bandwidth.
The loss is summed over the training characters (we omit this summation for clarity).
We note that this loss is differentiable wrt all the parameters of the joint prediction stage, including the bandwidth. The mean shift iterations of Eq. \ref{eq:mean-shift} are differentiable with respect to the attention weights and displaced points.
%\zhan{during training, we don't have non-maximum suppression, points are just gathered step by step. Here may need some modification. Besides, similar to \cite{kong2018grouppixels} we also have $L_{cd}$ after each meanshift step. We can mention that.} 
This allows us to backpropagate joint location error signal to both the vertex displacement and attention network. The Epanechnikov kernel in mean-shift is also a quadratic function wrt the bandwidth, which makes it possible to learn the bandwidth efficiently through  gradient descent. Learning converged to a value of $h=0.057$ based on our training dataset.

We also found that adding supervisory signal to the vertex displacements before clustering helped improving training speed and joint detection performance (see also experiments). To this end, we  minimize  Chamfer distance between  displaced points and  ground-truth joints,  favoring tighter clusters:
\begin{equation}
L_{cd}'(\bw_d) = \frac{1}{V}  \sum_{v} \min\limits_{k} || \bq_v - \hat{\bt}_k || +  \frac{1}{K}\sum_{k} \min\limits_{v} || \bq_v - \hat{\bt}_k ||
\label{eq:vertex_sup}
\end{equation}
%Here, we divide with  the number of mesh vertices to avoid sensitivity to mesh resolution. 
This loss affects only the parameters $\bw_d$ of the displacement module. Finally, we found that adding supervision to the vertex attention weights also offered a performance boost, as discussed in our experiments. This loss is driven by the observation that the displacement of vertices located closer to  joints are more helpful to localize them more accurately. Thus, for each training mesh, we find vertices closest to each joint at different directions perpendicular to the bones. Then we create a binary mask $\hat{\bm}$\ whose values are equal to $1$ for these closest vertices, and $0$ for the rest. We use cross-entropy to measure consistency between these masks and neural attention: 
 \begin{equation*}
 L_m(\bw_a) =  \hat{\bm}  \log \ba +(1-\hat \bm)\log(1-\ba) 
 \end{equation*}
 
 \vspace{-1mm}
\paragraph{Edge dropout.} During training of GMEdgeNet, for each batch, we randomly select a subset of edges within geodesic neighborhoods (in our implementation, we randomly select  subsets up to $15$ edges). This sampling strategy can be considered as  a form of mesh edge dropout. We found that it improved performance since it simulates varying vertex sampling on the mesh, making the graph network more robust to different tessellations. 

\begin{figure*}[t!]
  \centering
  \includegraphics[width=\linewidth]{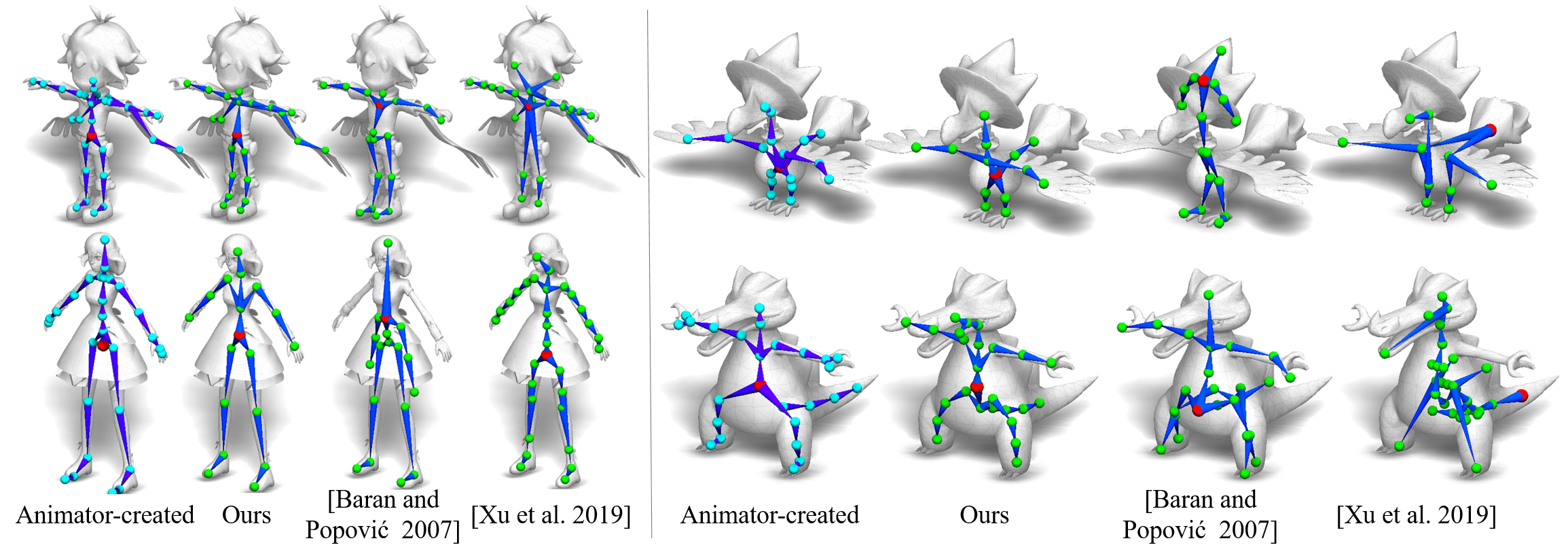}
  \vspace{-8mm}
  \caption{Comparisons with previous methods for skeleton extraction. For each character, the reference skeleton is shown on the left (``animator-created''). Our predictions tend to agree more with the reference skeletons. }
  \label{fig:comparison_skeleton}
  \vspace{-2mm}  
\end{figure*}

\begin{figure*}[h!]
  \centering
  \includegraphics[width=\linewidth]{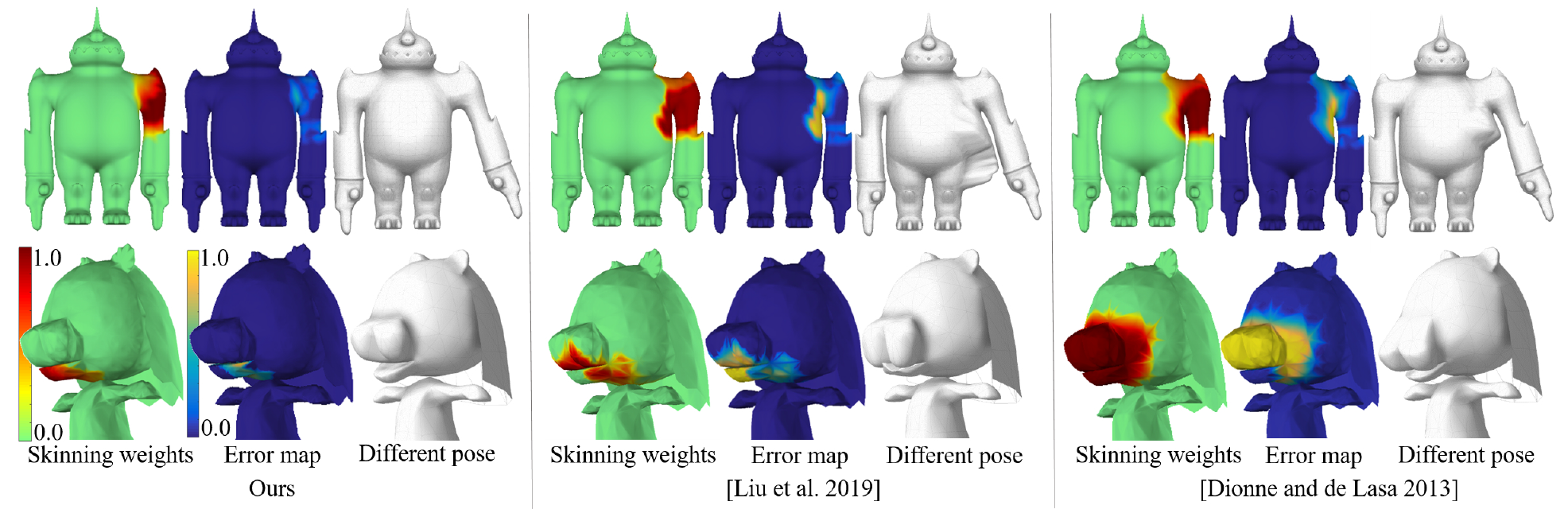}
  \vspace{-8mm}
  \caption{Comparisons with prior methods for skinning. We visualize skinning weights, L1 error maps, and a different pose (moving right arm for the robot above, and lowering the jaw of the character below). Our method produces lower errors in skinning weight predictions on average.}
  \label{fig:comparison_skinning}
  \vspace{-2mm}  
\end{figure*}

  \vspace{-1mm}
\paragraph{Training implementation details} We first pre-train the parameters $\bw_a$ of attention module with the loss $L_m$ alone. We found  that
bootstrapping the attention module with this pre-training helped with the performance (see also experiments). Then we fine-tune $\bw_a$, and train the parameters $\bw_d$ of the displacement module and the bandwidth $h$ using the combined loss: $L_{cd}(\bw_a, \bw_d, h)+L_{cd}'(\bw_d)$. For fine-tuning, we use the Adam optimizer with a batch size of $2$ training characters,
and learning rate $10^{-6}$.  

\subsection{Connectivity stage training} 
Given a training character, we form the adjacency matrix encoding the connectivity of the skeleton i.e., $\hat{p}_{ij}=1$ if two training joints $i$ and $j$ are connected, and $\hat{p}_{ij}=0$ otherwise
. The parameters $\bw_b$ of the BoneNet are learned using binary cross-entropy between the training adjacency matrix entries and the predicted probabilities $p_{i,j}$:
 \begin{equation*}
 L_m(\bw_a) =  \sum_{i,j}\hat{p}_{ij} \log p_{i,j} +(1-\hat{p}_{ij})\log(1-p_{i,j}) 
 \end{equation*}
The BoneNet parameters are learned using the probabilities $p_{i,j}$ estimated for training joints rather than the predicted ones of the previous stage. The reason is that the training adjacency matrix is defined on training joints (and not on the predicted ones). We  tried to find   correspondences between the predicted joints and the training ones using the Hungarian method, then transfer the training adjacencies  to pairs of matched joints. However, we did not observe  significant improvements by doing this potentially due to matching errors. Finally, to train the parameters $\bw_r$ of the network used to extract the root joint, we use the softmax loss for classification.

\vspace{-1mm}
\paragraph{Training implementation details.} Training BoneNet  has an additional challenge due to class imbalance problem: out of all pairs of joints, only few are connected. To deal with this issue, we adopt the online hard-example mining approach from \cite{shrivastava2016training}.  
For both networks, we  employ the Adam optimizer with batch size $12$ and learning rate $10^{-3}$.

\subsection{Skinning stage training} 
Given a set of training characters, each with skin weights $\hat{\bS}=\{\hat{\bs}_v\}$, we train the parameters $\bw_s$ of our skinning network so that the estimated skinning weights $\bS=\{\bs_v\}$ agree as much as possible with the training ones. By treating the per-vertex skinning weights as probability distributions, we  use cross-entropy as loss to quantify the disagreement between training and predicted distributions for each vertex:
 \begin{equation*}
 L_{s}(\bw_s) = \frac{1}{V} \sum_v \sum_{r} \hat{s}_{v,r} \log s_{v,r}
 \end{equation*}
 
As in the case of the connectivity stage,
we train the skinning network
based on the training skeleton
rather than the predicted one, since we do not have skinning weights for it.  We  tried to transfer skinning weights from the training bones to the predicted ones by establishing correspondences
as before, but this did not result in significant improvements.

\vspace{-1mm}
\paragraph{Training implementation details.} To train the skinning network, we  use the Adam optimizer with a batch size of $2$ training characters,
and learning rate $10^{-4}$. We also apply the edge dropout scheme during the training of this stage, as in the joint prediction stage.

%% file: chapters/results.tex
We evaluated our method and alternatives for  animation skeleton and skinning prediction both quantitatively and qualitatively. Below we discuss the dataset used for evaluation, the performance measures, comparisons, and ablation study. 

\vspace{-1mm}
\paragraph{Dataset.} To train and test our method and alternatives, we chose the ``ModelsResource-RigNetv1'' dataset of 3D\ articulated characters 
  from \cite{Xu19skeleton}, which provides a non-overlapping training and  test split,  and contains diverse  characters
  \footnote{please see also our project page:
  \textcolor{blue}{https://zhan-xu.github.io/rig-net}}.
Specifically, the dataset contains $2703$ rigged characters mined from an online repository \cite{ModelsResource}, spanning several categories, including humanoids, quadrupeds, birds, fish, robots, toys, and other fictional characters. Each character includes one rig (we note that the multiple rig examples of the two models of Figure \ref{fig:variance} were made separately and do not belong  to this dataset). 
The dataset does not contain duplicates, or re-meshed versions of the same character. Such duplicates were eliminated from the dataset. Specifically, all models were voxelized in a binary $88^3$ grid, then for each model in the dataset, we computed the Intersection over Union (IoU) with all other models based on their volumetric representation. We eliminated duplicates or near-duplicates whose IoU of volumes was more than 95\%. We also manually verified that such re-meshed versions were filtered out. Under the guidance of an artist, we also verified that all characters have plausible skinning weights and deformations.
We use a training, hold-out validation, and test split, following a $80\%$-$10\%$-$10\%$ proportion respectively, resulting in $2163$ training, $270$ hold-out validation, and $270$ test characters.
Figure \ref{fig:training} shows examples from the training split.   
 The models are consistently oriented and scaled. Meshes with fewer than $1K$ vertices were subdivided; as a result all training and test meshes contained between $1$K and $5$K vertices.
 The  number of joints per character varied from $3$ to $48$, and the average is $25.0$. The quantitative and qualitative evaluation was performed on the test split of the dataset.

\begin{figure*}[t!]
  \centering
  \includegraphics[width=\linewidth]{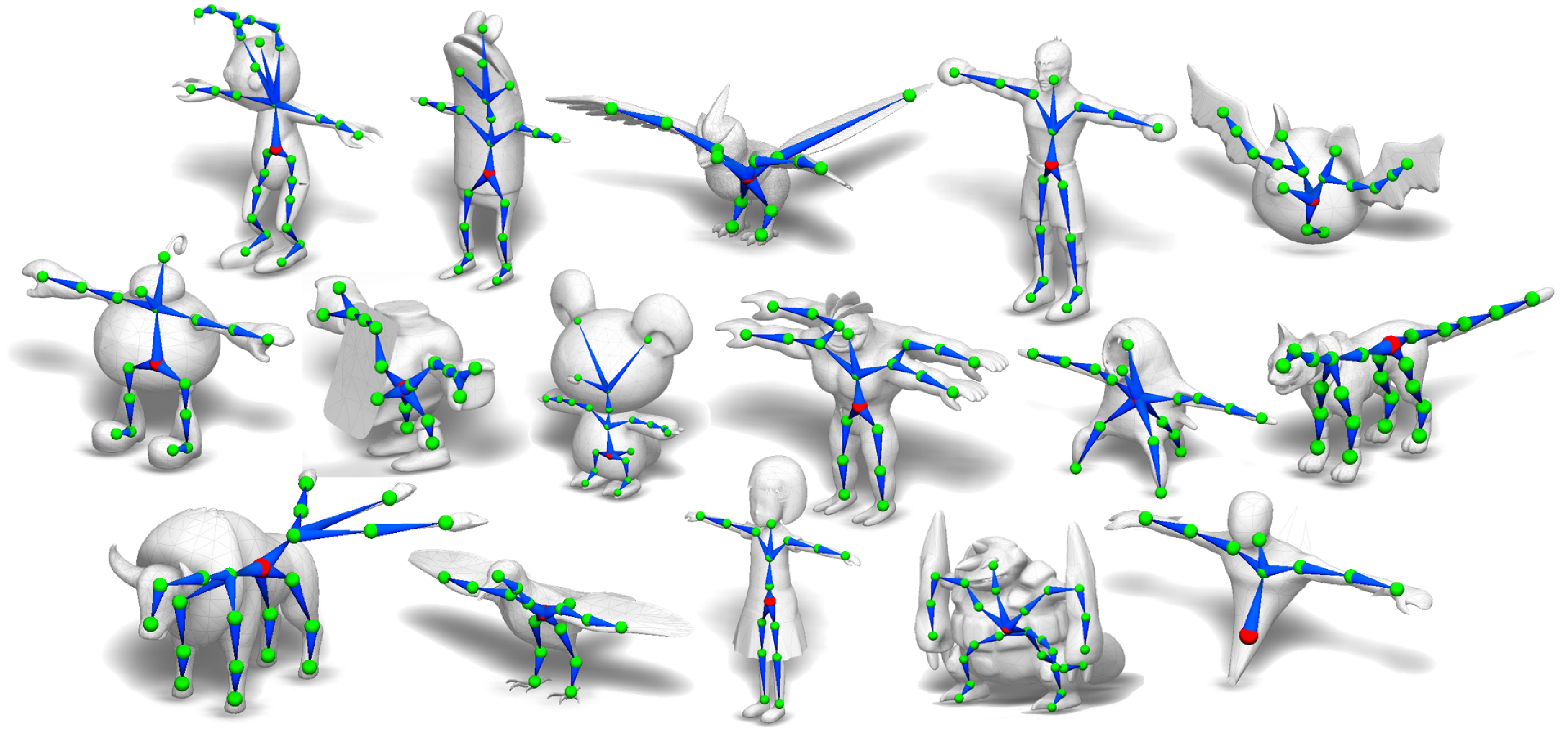}
  \vspace{-7mm}
  \caption{Predicted skeletons for  test models with varying structure and morphology.
  Our method is able to produce reasonable skeletons even for models that have different number or types of parts than the ones used for training e.g., quadrupeds with three tails.
}
  \label{fig:gallery}
  \vspace{-2mm}  
\end{figure*}

\begin{wrapfigure}{R}{0.5\linewidth}
 \vspace{-2mm} 
  \includegraphics[width=1\linewidth]{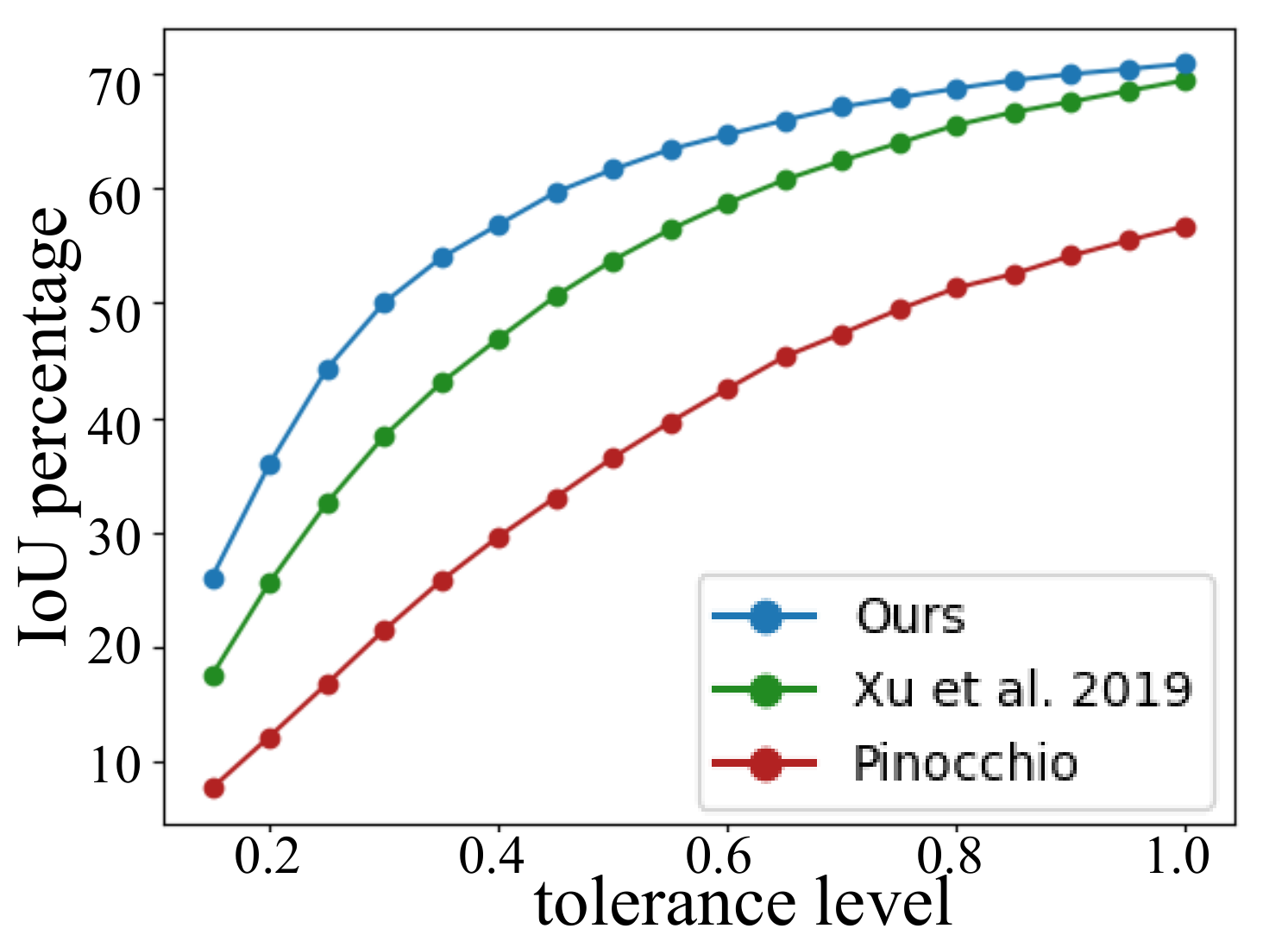}
  \vspace{-6mm}
  \caption{IoU vs different tolerances.}
\vspace{-4mm}     
  \label{fig:iou_curve}
\end{wrapfigure}

\vspace{-1mm}
\paragraph{Quantitative evaluation measures.} Our quantitative evaluation aims to  measure the similarity of the predicted animation skeletons and skinning  to the ones created by modelers in the test set (denoted as ``reference skeletons'' and ``reference skinning'' in the following paragraphs).
For evaluating skeleton similarity, we employ various measures  following   \cite{Xu19skeleton}: \\
(a) \emph{CD-J2J} is the symmetric Chamfer distance between joints. Given a test shape, we measure the Euclidean distance from each predicted joint to the nearest joint in its reference skeleton, then divide with the number of predicted joints.  We also compute the Chamfer distance the other way around from the reference skeletal joints to the nearest predicted ones. We denote the average of the two  as \emph{CD-J2J}.\\ 
(b) \emph{CD-J2B} is the Chamfer distance between joints and bones. The difference from the previous measure is that  for each predicted joint, we compute its distance to the nearest bone point on the reference skeleton.  We symmetrize this measure by also computing the distance from  reference  joints to predicted bones. A low value of \emph{CD-J2B} and a high value of \emph{CD-J2J} mean that the predicted and reference skeletons tend to overlap, yet the joints are misplaced along the bone direction.\\
(c) \emph{CD-B2B} is the Chamfer distance between bones (line segments). As above, we define it symmetrically. CD-B2B measures similarity of skeletons in terms of bone placement (rather than joints). Ideally, all \emph{CD-J2J},  \emph{CD-J2B}, and \emph{CD-B2B}  measures should be low. \\
(d) \emph{IoU}  (Intersection over Union) can also be used to characterize  skeleton similarity. First, we find a maximal matching between the predicted and reference joints by using the Hungarian algorithm. Then we measure the number of predicted and reference joints that are matched and whose Euclidean distance is lower then a prescribed tolerance. This is then divided with the total number of predicted and reference joints.
By varying the tolerance, we can obtain plots demonstrating IoU for various tolerance levels (see Figure \ref{fig:iou_curve}). 
To  provide a single, informative value, we set the tolerance to half of the  local shape diameter \cite{Shapira:2008:CMP} evaluated at each corresponding reference joint. This is evaluated by casting rays perpendicular to the bones  connected at the reference joint, finding ray-surface intersections, and computing the joint-surface distance averaged over all  rays. The reason for this normalization is that thinner parts e.g, arms have lower shape diameter; as a result, small joint deviations can cause more noticeable misplacement compared to thicker parts like torso.\\
(e) \emph{Precision \& Recall} can also be used here.
 Precision is the  fraction of predicted joints that were matched and whose distance to their nearest reference one is lower than the tolerance defined above.
Recall is the fraction of  reference joints that were matched and whose distance to  their nearest predicted joints is lower than the tolerance. 
Note that since the number of reference or predicted joints may not be the same. Unmatched predicted joints contribute no precision, and similarly unmatched reference joints contribute no  recall.\\
(f) \emph{TreeEditDist (ED)} is the tree edit distance measuring the topological difference of the predicted skeleton to the reference one. The measure is defined as the minimum number of joint deletions, insertions, and replacements that are necessary to transform the predicted skeleton into the reference one.

\begin{table}[t!]
\centering % used for centering table
\begin{tabular}{|@{}c@{}| @{}c@{}| @{}c@{}| @{}c@{}| @{}c@{}|@{}c@{}|@{}c@{}|@{}c@{}| } 
\hline %inserts double horizontal lines
  & \,IoU\, & \,Prec.\, & \,Rec. \, & \,CD-J2J\, & \,CD-J2B\, & \,CD-B2B\,  \\ %& \,ED\, \\ [0.5ex]
%heading
\hline % inserts single horizontal line
\hline % inserts single horizontal line

Pinocchio & 36.5\%\, & \,38.7\%\, & \,35.9\% &7.2\% &5.5\% & 4.7\% \, \\ % & 12.5\\
Xu et al. 2019 & 53.7\%& 53.9\% & 55.2\% &4.5\% &2.9\%& 2.6\% \, \\ % 13.2\\
\hline
Ours & \textbf{61.6\%} & \textbf{67.6\%}& \textbf{58.9\%} & \textbf{3.9\%} & \textbf{2.4\%} & \textbf{2.2\%} \, \\
% & \textbf{12.1}\\  % [1ex] adds vertical space
\hline %inserts single line
\end{tabular}
\caption{ Comparisons with other skeleton prediction methods. } % title of Table
\vspace{-8mm}
\label{table:joint_comparison} % is used to refer this table in the text
\end{table}

To evaluate skinning, we use the reference skeletons for all methods, and measure similarity between predicted and reference skinning maps:\\ (a) \emph{Precision \&\ Recall} are measured by  finding the set of bones that influence each vertex significantly, where influence corresponds to a skinning weight larger than a threshold ($1e^{-4}$, as  described in \cite{Liu2019}). Precision is the fraction of influential bones based on the predicted skinning among the ones defined based on the reference skinning. Recall is the fraction of the influential bones based on the reference skinning matching the ones found from the predicted skinning. \\
(b) \emph{L1-norm} measures the L1 norm of the difference between the predicted skinning weight vector and the reference one for each mesh vertex. We compute the average L1-norm over each test mesh. \\
(c) \emph{dist} measures the Euclidean distance between the position of vertices deformed based on the reference skinning and the predicted one. To this end, given a test shape, we generate $10$ different random poses,  and compute  the average and max distance error over the mesh vertices.

All the above skeleton and skinning evaluation measures are computed for each test shape, then averaged over the the test split.

\begin{table}[t!]
\centering % used for centering table
\begin{tabular}{|@{}c@{}| @{}c@{}| @{}c@{} | @{}c@{} |@{}c@{}|@{}c@{}| }
\hline %inserts double horizontal lines
  & \,Prec.\, & \,Rec.\,  & \,avg L1\, &  \,avg dist\, & \,max dist \, \\ [0.5ex]
%heading
\hline % inserts single horizontal line
\hline % inserts single horizontal line
BBW        & \,68.3\%\, & \,77.6 \%\, & 0.69 &0.0061 &0.055\\
GeoVoxel   & \,72.8\%\, & \,75.1 \%\, & 0.65 &0.0057 &0.049\\
NeuroSkinning  &76.3\% &74.7 \%&0.57 &0.0053 &0.043 \\
\hline % inserts single horizontal line
Ours & \textbf{82.3\%} & \textbf{80.8\%} & \textbf{0.39} & \textbf{0.0041}& \textbf{0.032}\\  % [1ex] adds vertical space
\hline %inserts single line
\end{tabular}
\caption{ Comparisons with other skinning prediction methods.} % title of Table
\label{table:skinning_comparison} % is used to refer this table in the text
\vspace{-5mm}
\end{table}

\vspace{-1mm}
\paragraph{Competing methods.} For skeleton prediction, we compare our method with \emph{Pinocchio} \cite{Baran:2007:ARA} and \cite{Xu19skeleton}. Pinocchio fits a template skeleton for each model. The template is automatically selected among a set of predefined ones (humanoid, short quadruped, tall quadruped, and centaur) by evaluating the fitting cost for each of them, and choosing the one with the least cost. %Pinocchio also provides a skinning method based on heat equilibrium - we use its default parameters \kalo{explain}.
\cite{Xu19skeleton} is a learning method trained on the same split as ours, with hyper-parameters tuned in the same validation split.
%The method does not perform skinning.  
For skinning weights prediction, we compare with the Bounded-Biharmonic Weights (BBW) method \cite{Jacobson11}, NeuroSkinning \cite{Liu2019} and the geometric method from \cite{Dionne13}, called ``GeoVoxel''. For the BBW method, we adopt the implementation from libigl \cite{libigl}, where the mesh is first tetrahedralized, then the bounded biharmonic weights are computed based on this volume discretization. For NeuroSkinning, we  trained the network on the same split as ours and optimized its hyperparameters in the same hold-out validation split. For GeoVoxel, we adopt Maya's implementation \cite{Maya} which outputs skinning weights based on a hand-engineered function of volumetric geodesic distances. We set the max influencing bone number, weight pruning threshold, and drop-off parameter through holdout validation in our validation split (3 bones, 0.3 pruning threshold, and 0.5 dropoff).

\setlength{\columnsep}{10pt}
\vspace{-1mm}
\paragraph{Comparisons.} Table \ref{table:joint_comparison} reports the evaluation measures for skeleton extraction between competing techniques.  Our method  outperforms the rest according to all measures. This is also shown in Fig.\ref{fig:iou_curve}, showing IoU on the y-axis for different tolerance levels (multipliers of local shape diameter) on the x-axis.

\begin{table}[t!]
\centering % used for centering table
\begin{tabular}{|@{}c@{}| @{}c@{}| @{}c@{}| @{}c@{}| @{}c@{}|@{}c@{}|@{}c@{}|@{}c@{}| } 
\hline %inserts double horizontal lines
   & \,IoU\, & \,Prec.\, & \,Rec. \, & \,CD-J2J\, & \,CD-J2B\,  & \,CD-B2B\,\\ [0.5ex]
%heading
\hline % inserts single horizontal line
\hline % inserts single horizontal line
P2PNet-based & 40.6\% & 41.6\%& 42.0\% &6.3\% &4.6\%& 3.8\% \\
No attn  & 52.4\%& 50.9\%& 50.7\% &4.6\% & 3.1\% & 2.7\%\\
One-ring &59.7\%&65.6\%&57.4\% &4.1\% &2.5\%&2.4\%\\
No vertex loss  & 59.3\%& 58.2\%& 57.6\% &4.2\% &2.7\%& 2.5\%\\
No attn pretrain  &60.6\%&64.0\%&58.1\%&4.2\% &2.6\%&2.4\%\\
\hline % inserts single horizontal line

Full & \textbf{61.6\%} & \textbf{67.6\%}& \textbf{58.9\%} & \textbf{3.9\%} & \textbf{2.4\%} & \textbf{2.2\%} \\  % [1ex] adds vertical space
\hline %inserts single line
\end{tabular}
\caption{Joint prediction ablation study} % 
\label{tab:joint_ablation} % is used to refer this table in the text
\vspace{-7mm}
\end{table}

Figure \ref{fig:comparison_skeleton} visualizes  reference skeletons and  predicted ones for different methods for some characteristic test shapes. We observe that our method tends to output skeletons whose joints and bones are closer to the reference ones. \cite{Baran:2007:ARA} often produces implausible skeletons when the input model has parts (e.g., tail, clothing) that do not correspond well to the used template. \cite{Xu19skeleton} tends to misplace joints around  areas, such as elbows and knees,  since\ voxel grids tend to lose surface detail.

Table \ref{table:skinning_comparison} reports the evaluation measures for skinning. Our numerical results are significantly better than BBW, NeuroSkinning, and GeoVoxel according to all the measures. Figure \ref{fig:comparison_skinning} visualizes the skinning weights produced by our method, GeoVoxel, and NeuroSkining that were found to be the best alternatives according to our numerical evaluation. Ours tends to agree more with the artist-specified skinning. On the top example, arms are close to torso in terms of Euclidean distance, and to some degree also in geodesic sense. Both NeuroSkining and GeoVoxel  over-extend the skinning weights to a  larger area than the arm. In order to match the GeoVoxel's output to the artist-created one, all its parameters need to be manually tuned per test shape, which is laborious. Our method combines  bone representations and vertex-skeleton intrinsic distances in our mesh network to produce skinning that better separates articulating parts. In the bottom example, a jaw joint is placed close to the lower lip  to control the jaw animation. Most vertices on the front face are close to this joint in terms of both geodesic and Euclidean distances. This results in higher errors for both NeuroSkinning and GeoVoxel, even if the latter is manually tuned. Our method produces a sharper map  capturing the part of the jaw.

\begin{table}[t!]
\centering
\begin{tabular}{|c|c|c|c|}
\hline %inserts double horizontal lines
       & Class. Acc. & CD-B2B & ED \\
\hline
\hline
  Euclidean edge cost            & 61.2\% & 0.30\% & 5.0 \\
  bone descriptor only           & 71.9\% & 0.22\% & 4.2 \\ 
  bone descriptor+skel. geometry & 80.7\% & 0.12\% & 2.9 \\
  \hline
  Full stage & \textbf{83.7\%} & \textbf{0.10\%} & \textbf{2.4} \\
\hline
\end{tabular}
\caption{Connectivity prediction ablation study} % title of Table
\label{tab:connectivity_ablition} % is used to refer this table in the text
\vspace{-5mm}
\end{table}

\begin{table}[t!]
\centering % used for centering table
\begin{tabular}{|@{}c@{}| @{}c@{}| @{}c@{}| @{}c@{}| @{}c@{}|@{}c@{}|@{}c@{}| }
\hline %inserts double horizontal lines
  & \,Prec\, & \,Rec.\,  & \,avg-L1\, & \,avg-dist.\, & \,max-dist. \, \\ [0.5ex]
%heading
\hline % inserts single horizontal line
\hline % inserts single horizontal line
No geod. dist. &80.0\% &79.3\%& 0.41 &0.0044 &0.054 \\
\hline % inserts single horizontal line
Ours & \textbf{82.3\%} & \textbf{80.8\%} & \textbf{0.39} & \textbf{0.0041}& \textbf{0.032}\\

  % [1ex] adds vertical space
\hline %inserts single line
\end{tabular}
\caption{Skinning prediction ablation study} % title of Table
\label{tab:skinning_ablation} % is used to refer this table in the text
\vspace{-5mm}
\end{table}

\vspace{-1mm}
\paragraph{Ablation study.} We present the following ablation studies to demonstrate the influence from different design choices of our method.\\
(a) \emph{Joint prediction ablation study}: Table \ref{tab:joint_ablation} presents evaluation of  variants of our joint detection stage  trained in the same split and tuned in the same hold-out validation split as our original method.  We examined the following variants: \textit{``P2PNet-based''} uses the same architecture as P2PNet \cite{yin2018p2pnet}, which relies on PointNet \cite{qi2017pointnetplusplus}
for displacing points (vertices in our case). After displacement, mean-shift clustering is used to extract joints as in our method. We experimented with the loss from their approach, and also the same loss as in our joint detection stage (excluding the attention mask loss, since P2PNet does not use attention). The latter choice worked better. The architecture was trained and tuned in the same split as ours.
 \textit{``No attn''} is our method without the attention module, thus all vertices have the same weight during clustering. \textit{``One-ring''} is our method where GMEdgeConv uses only one-ring neighbors of each vertex without considering geodesic neighborhoods. \textit{``No vertex loss''} does not use vertex displacement supervision with the Chamfer distance loss of Eq. \ref{eq:vertex_sup} during training. It uses supervision from clustering only based on the loss of Eq.\ref{eq:cluster_sup}. \textit{``No attn pretrain''} does not pre-train the attention network with our created binary mask. We observe that removing any of these components, or using an architecture based on P2PNet, leads to a noticeable performance drop. In particularly, the attention module has a significant influence on the performance of our method.\\
(b) \emph{Connectivity prediction ablation study.} Table \ref{tab:connectivity_ablition} presents evaluation of alternative choices for our BoneNet. In these experiments, we examine the performance of the connectivity module when it is given as input
the  reference joints instead of the predicted ones. In this manner, we specifically evaluate the design choices for the connectivity stage i.e., our evaluation here is not affected from any wrong predictions of the joint detection stage. Here, we report the binary classification accuracy (``Class. Acc.'') i.e., whether the prediction to connect each pair of given joints agrees with the ground-truth connectivity. We also report edit distance (ED) and bone-to-bone Chamfer distance (CD-B2B), since these measures are specific to bone evaluation. We first show the performance when the MST connects  joints based on  Euclidean distance as cost (see ``Euclidean edge cost''). We  also evaluate the effect
of using only the bone
descriptor
without 
 the skeleton geometry encoding ($\bg_t$) and without shape encoding ($\bg_s$) (see ``bone descriptor only'', and Eq.\ref{eq:bonenet}). We also evaluate the effect
of using  the bone
descriptor
with 
 the skeleton geometry encoding but without shape encoding  (see ``bone descriptor+skel. geometry''). The best performance is achieved when all three shape, skeleton, and bone representations  are used as input to BoneNet.
We also observed the same trend in  RootNet, where we evaluate the accuracy of predicting the  root joint correctly. Skipping the skeleton geometry and shape encoding results in  accuracy of $67.8\%$. Adding the skeleton encoding increases it to $86.8\%$. Using  all three shape, skeleton, and joint representations achieves the best accuracy of $88.9\%$. \\

\setlength{\columnsep}{8pt}
\begin{wrapfigure}{R}{0.5\linewidth}
 \vspace{-4mm} 
  \includegraphics[width=1\linewidth]{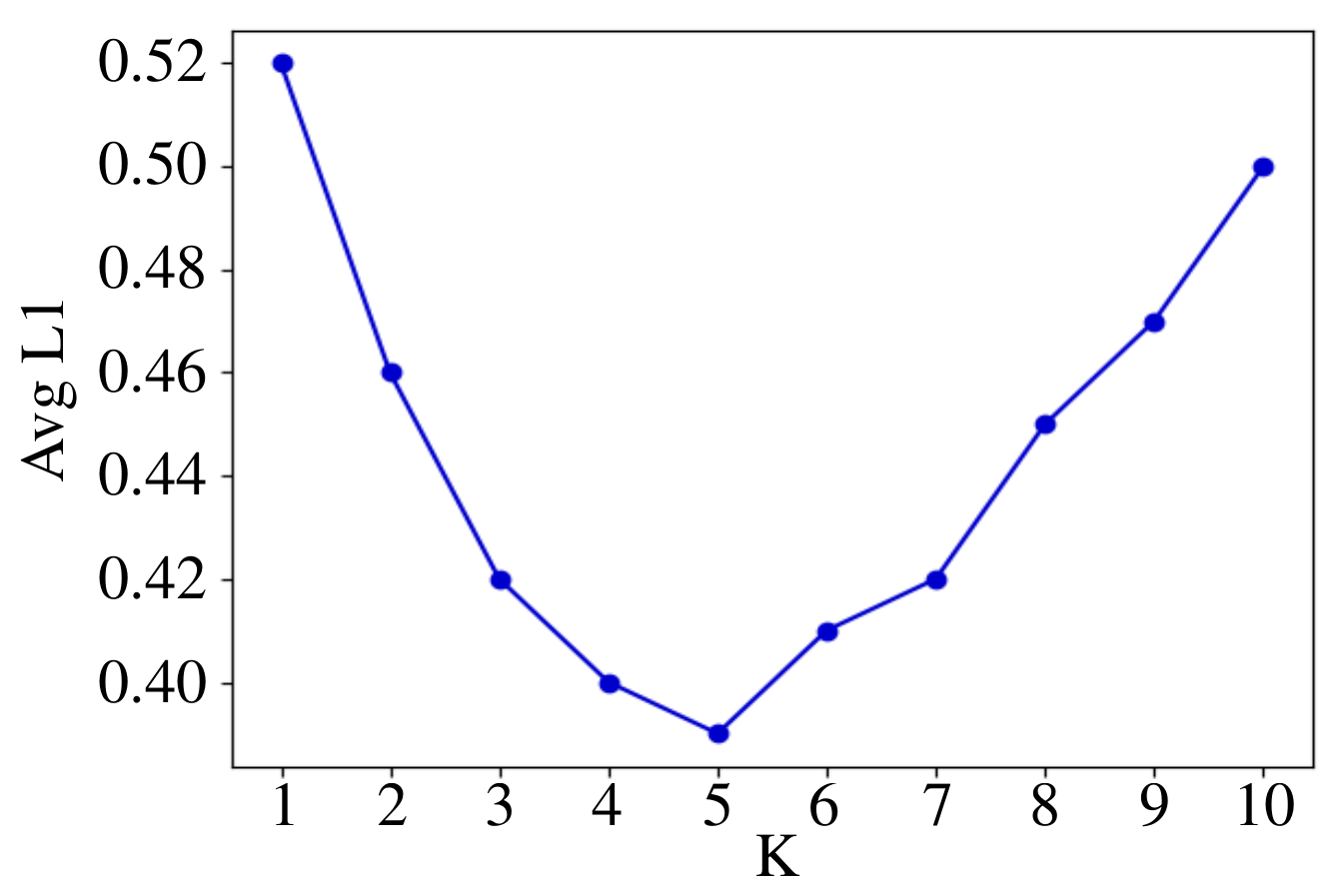}
  \vspace{-9mm}
  \caption{Skinning weight error wrt different number $K$ of closest bones used in our network.}
\vspace{-4mm}     
  \label{fig:different_k}
\end{wrapfigure}

\vspace{-1mm}
(c) \emph{Skinning prediction ablation study}. Table \ref{tab:skinning_ablation} presents the case of removing the volumetric geodesic distance feature from input to our skinning prediction network. We observe a noticeable performance drop without it. Still, it is interesting to see that even without it,  our method is better than competing methods (Table \ref{table:skinning_comparison}). We also experimented with different choices of $K$ i.e., the number of closest bones used in our skinning prediction. Fig.\ref{fig:different_k} shows the average L1-norm difference of skinning weights for $K=1...10$ in our test set. Lowest error is achieved when $K=5$ (we noticed the same behavior and minimum in our validation split).

%% file: chapters/conclusion.tex
We presented a method that automatically rigs input 3D character models. To the best of our knowledge, our method represents a first step towards a learning-based, complete solution to character rigging, including skeleton creation and skin weight prediction. 
We believe that our method is practical in various scenarios. First, we believe that our method is useful for casual users or novices, who might not have the training or expertise to deal with modeling and rigging interfaces. Another motivation for using our method is the widespread effort for democratization of 3D content creation and animation that we currently observe in online asset libraries provided with modern game engines (e.g., Unity). We see our approach as such one step towards further democratization of character animation. Another scenario of use for our method is when a large collection of 3D characters need to be rigged. Processing every single model manually would be cumbersome even for experienced artists.

\setlength{\columnsep}{7pt}
\begin{wrapfigure}{R}{0.36\linewidth}
 \vspace{-4mm} 
  \includegraphics[width=0.98\linewidth]{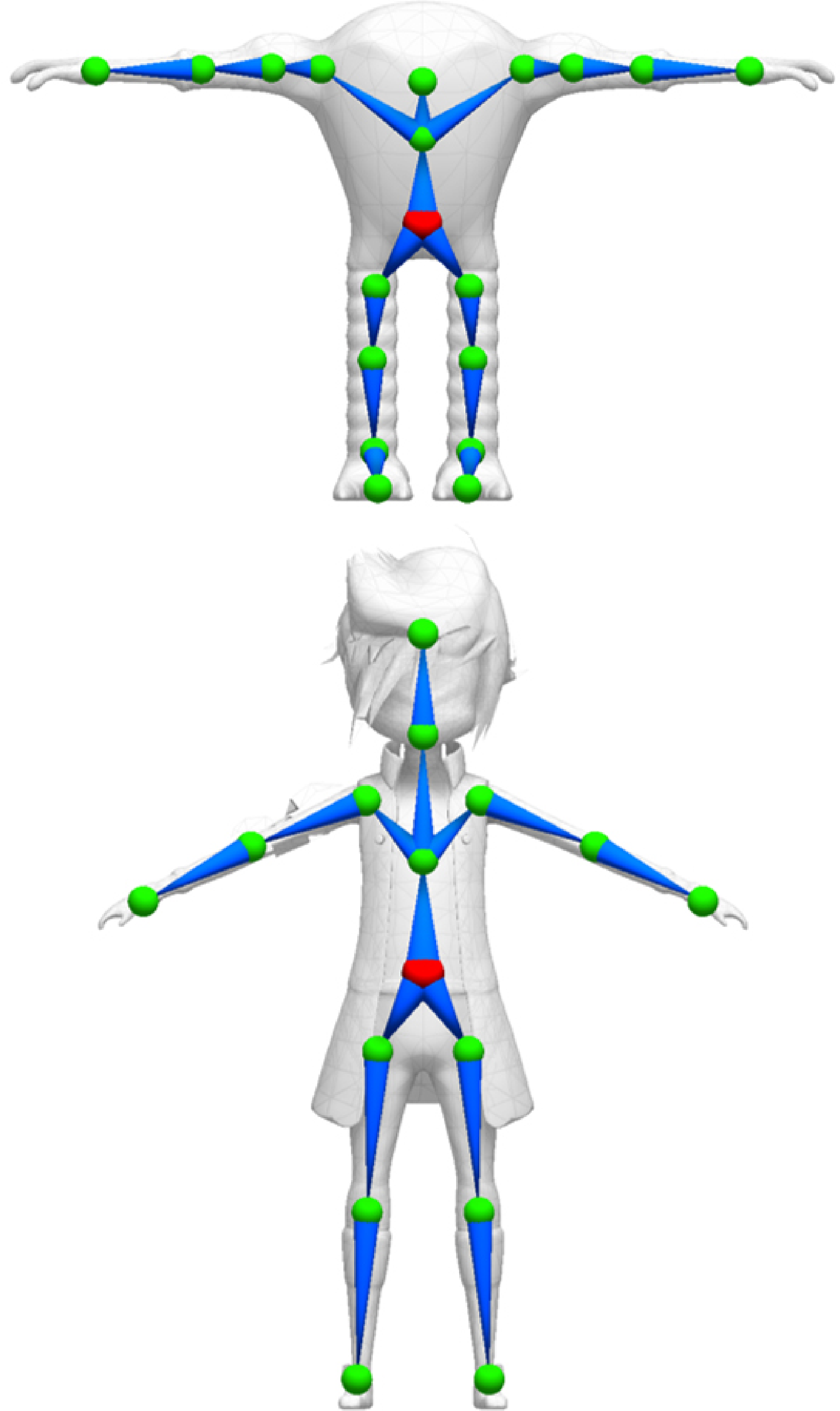}
  \vspace{-3mm}
  \caption{
  Failure cases. \emph{(Top:)} extra joints in the arms. \emph{(Bottom:)} missing helper joints for clothes.}
\vspace{-2mm}     
  \label{fig:limitations}
\end{wrapfigure}
Our approach does have limitations, and exciting avenues for future work. First, our method currently uses a per-stage training approach. Ideally, the skinning  loss could be back-propagated to all stages of the network to improve joint prediction. However, this implies
differentiating volumetric geodesic distances and skeletal structure estimation, which are hard tasks. Although we trained our method such that it is more robust to different vertex sampling and tessellations, invariance to mesh resolution and connectivity is not guaranteed.  Investigating the performance of other mesh neural networks (e.g., spectral) here, could be impactful. 
There are few cases where our method produces undesirable effects, such as putting extra arm joints (Figure \ref{fig:limitations}, top).
Our dataset also has limitations.  It contains one rig per model. Many rigs often do not include bones for small parts, like feet, fingers, clothing and accessories, which makes our trained model less predictive of these joints
(Figure \ref{fig:limitations}, bottom).
Enriching the dataset with more rigs could improve performance, though it might make the mapping  more multi-modal than it is at present.  A multi-resolution approach that refines the skeleton in a coarse-to-fine manner may instead be fruitful. Our current bandwidth parameter explores one mode of variation. Exploring a richer space to interactively control skeletal morphology and resolution is another interesting research direction. Finally, it would also be interesting to extend our method to handle skeleton extraction for point cloud recognition or reconstruction tasks.

%% file: chapters/suppl.tex
Table \ref{tab:layers} lists the layer used in each stage of our architecture along with the size of its output map. We also note that our project page with source code, datasets, and supplementary video is available at:\\ \textcolor{blue}{https://zhan-xu.github.io/rig-net}.

\begin{table}[!b]
\begin{tabular}{|@{}c@{}|@{}c@{}|@{}c@{}|}
\hline
\multicolumn{3}{|c|}{Joint Prediction Stage}                       \\ \hline
Layers                            & Input         & Output         \\ \hline
GMEdgeConv                        & $V\times3\ (x\_0)$    & $V\times64\ (x\_1)$          \\ \hline
GMEdgeConv                        & $V\times64$          & $V\times256\ (x\_2)$         \\ \hline
GMEdgeConv                        & $V\times256$         & $V\times512\ (x\_3)$         \\ \hline
concat($x\_1, x\_2, x\_3$)        &               & $V\times832$                \\ \hline
MLP ({[}832, 1024{]})             & $V\times832$         & $V\times1024$               \\ \hline
max\_pooling \& tilt              & $V\times1024$        & $V\times1024\ (x\_glb)$      \\ \hline
concat($x\_0, x\_1, x\_2, x\_3, x\_glb$)    &               & $V\times1859$               \\ \hline
MLP({[}1859, 1024, 256, 3{]})           & $V\times1859$        & $V\times3$                  \\ \hline
\multicolumn{3}{|c|}{Connectivity Stage}                                            \\ \hline
GMEdgeConv                        & $V\times3\ (x\_0)$    & $V\times64\ (x\_1)$          \\ \hline
GMEdgeConv                        & $V\times64$          & $V\times128\ (x\_2)$         \\ \hline
GMEdgeConv                        & $V\times128$         & $V\times256\ (x\_3)$         \\ \hline
concat($x\_1, x\_2, x\_3$)                &               & $V\times448$                \\ \hline
MLP ({[}448, 512, 256, 128{]})     & $V\times448$         & $V\times128$                 \\ \hline
max\_pooling \& tile               & $V\times128$          & $P\times128\ (g\_s)$        \\ \hline
MLP({[}3, 64, 128, 1024{]})             & $K\times3$           & $K\times1024$                \\ \hline
max\_pooling \& tilt               & $K\times1024$         & $P\times1024$                \\ \hline
MLP({[}1024, 256, 128{]})    & $P\times1024$  & $P\times128\ (g\_t)$          \\ \hline
MLP({[}8, 32, 64, 128, 256{]}))             & $P\times8$   & $P\times 256\ (f\_ij)$ \\ \hline
 concat($g\_s, g\_t, f\_ij$)        &   & $P\times 512$        \\ \hline
MLP({[}512, 128, 32, 1{]})             & $P\times512$ & $P\times1$          \\ \hline
\multicolumn{3}{|c|}{Skinning Stage}                                        \\ \hline
MLP([38, 128, 64])                & $V\times38$          & $V\times64\ (x\_0)$          \\ \hline
GMEdgeConv                        & $V\times64$          & $V\times512\ (x\_1)$         \\ \hline
max\_pooling \& tilt              & $V\times512$         & $V\times512$                \\ \hline
MLP({[}512, 512, 1024{]})         & $V\times512$         & $V\times1024\ (x\_glb)$      \\ \hline
GMEdgeConv                        & $V\times512\ (x\_1)$  & $V\times256\ (x\_2)$         \\ \hline
GMEdgeConv                        & $V\times256\ (x\_2)$  & $V\times256\ (x\_3)$         \\ \hline
concat($x\_glb, x\_3$)              &               & $V\times1280$               \\ \hline
MLP({[}1280, 1024, 512, 5{]})     & $V\times1280$        & $V\times5$                  \\ \hline
\end{tabular}
\caption{RigNet architecture details. $V$ is the number of vertices from the input mesh. $K$ is the number of predicted joints. $P$ is the number of candidate bones defined by all pairs of predicted joints.} % 
\label{tab:layers} 
\end{table}